\documentclass[12pt]{JHEP3}
\usepackage{amsfonts}
\usepackage{amssymb}
\usepackage{lscape}
\usepackage{cite}
\usepackage{epsfig}
\usepackage{multirow}
\usepackage{longtable}
\usepackage[all]{xy}
\usepackage{amsmath}
\author{}

\newcommand{\drawsquare}[2]{\hbox{%
\rule{#2pt}{#1pt}\hskip-#2pt
\rule{#1pt}{#2pt}\hskip-#1pt
\rule[#1pt]{#1pt}{#2pt}}\rule[#1pt]{#2pt}{#2pt}\hskip-#2pt
\rule{#2pt}{#1pt}}
\newcommand{\fund}{\raisebox{-.5pt}{\drawsquare{6.5}{0.4}}}
\newcommand{\Ysymm}{\raisebox{-.5pt}{\drawsquare{6.5}{0.4}}\hskip-0.4pt%
         \raisebox{-.5pt}{\drawsquare{6.5}{0.4}}}
\newcommand{\Yasymm}{\raisebox{-3.5pt}{\drawsquare{6.5}{0.4}}\hskip-6.9pt%
        \raisebox{3pt}{\drawsquare{6.5}{0.4}}}

\newcommand{\be}{\begin{equation}}
\newcommand{\ee}{\end{equation}}
\newcommand{\ba}{\begin{array}}
\newcommand{\ea}{\end{array}}
\newcommand{\bea}{\begin{eqnarray}}
\newcommand{\eea}{\end{eqnarray}}

\newcommand{\ov}{\overline}
\def\IR{\relax{\rm I\kern-.18em R}}

\def\IP{\relax{\rm I\kern-.18em P}}
\def\inbar{\vrule height1.5ex width.4pt depth0pt}
\def\IC{\relax\,\hbox{$\inbar\kern-.3em{\rm C}$}}

\def\K3{{\bf K3}}

\def\ov{\overline}

\def\n2d{\cN_{V^*}^{\otimes 2}}

\def\IC{\mathbb{C}}

\def\IR{\mathbb{R}}

\def\IP{\mathbb{P}}

\def\cN{{\mathcal N}}

\title{SU(5) D-brane realizations, Yukawa couplings and proton stability}
\author{
P. Anastasopoulos$^{1}$\footnote{pascal@hep.itp.tuwien.ac.at},~
G. K. Leontaris$^{2}$\footnote{leonta@uoi.gr},~
R. Richter$^{3}$\footnote{rrichter@roma2.infn.it}~,
A. N. Schellekens$^{4,5,6}$\footnote{t58@nikhef.nl}~\\
$^1$ Technische Univ. Wien Inst. f\"ur Theoretische Physik, A-1040 Vienna, Austria\\
$^2$ Theoretical Physics Division, Ioannina University, GR-45110 Ioannina, Greece\\
$^3$ I.N.F.N.\ -\ Sezione di Roma ``Tor Vergata'', 00133, Roma, Italy\\
$^4$ NIKHEF, Science Park 105, 1098 XG Amsterdam, The Netherlands\\
$^5$ IMAPP, Radboud Universiteit Nijmegen, The Netherlands\\
$^6$Instituto de F\'\i sica Fundamental, CSIC, Madrid, Spain}
\date{}

\maketitle
\abstract{
We discuss SU(5) Grand Unified Theories in the context of orientifold compactifications. Specifically, we investigate two and three D-brane stack  realizations of the Georgi-Glashow and the flipped SU(5) model and analyze them with respect to their Yukawa couplings. As pointed out in \cite{Kiritsis:2009sf} the most economical Georgi-Glashow realization based on two stacks generically suffers from a disastrous large proton decay rate. We show that allowing for an additional $U(1)$ D-brane stack this as well as other phenomenological problems can be resolved. We exemplify with globally consistent Georgi-Glashow models based on RCFT that these D-brane quivers can be indeed embedded in a global setting. These globally consistent realizations  admit  rigid O(1) instantons inducing the perturbatively missing coupling $ \mathbf{10}\mathbf{10}\mathbf{5^H}$. Finally we show that flipped SU(5) D-brane realizations even with multiple $U(1)$ D-brane stacks are plagued by severe phenomenological drawbacks which generically cannot be overcome.  
}
\preprint{
TUW-10-16\\
ROM2F/2010/18\\
NIKHEF/2010-038}
 \clearpage
\begin{document}

\section{Introduction}
D-brane compactifications have been proven to be a promising framework for realistic string model building. The basic building blocks of such constructions are D-branes which fill out the four-dimensional space-time and wrap submanifolds in the internal manifold.
The gauge bosons live on the world volume of the respective D-brane while chiral matter appears at intersections of different stacks of D-branes. The multiplicity of the latter is given by the number of intersections of the respective submanifolds in the internal space. Over the last decade many globally consistent semi-realistic D-brane models have been constructed  (for recent reviews, see \cite{Blumenhagen:2005mu,Blumenhagen:2006ci,Marchesano:2007de}) . 

In this work we are investigating how supersymmetric $SU(5)$ GUT's can be realized in this framework with the emphasis on the realization of the superpotential\footnote{For global supersymmetric $SU(5)$ D-brane realizations, see \cite{Cvetic:2002pj,Chen:2005aba,Chen:2005mm,Chen:2005cf,Gmeiner:2006vb,Floratos:2006hs,Cvetic:2006by,Antoniadis:2007jq, Blumenhagen:2008zz}. For a related study of GUT's within this framework, see \cite{Berenstein:2006aj}.}. The $SU(5)$ gauge symmetry arises from a stack $a$ of $5$ D-branes giving rise to the gauge symmetry $U(5)_a$ which further splits into $U(5)_a = SU(5) \times U(1)_a$. Here the abelian part generically acquires a mass via the Green-Schwarz mechanism and survives only as global symmetry. 
The $\mathbf{10}$ is localized at intersections of the $U(5)$ stack $a$ and its orientifold image $a'$. To accommodate for the other matter fields $\mathbf{\ov 5}$, as well as the Higgs pair $\mathbf{5^H}$ and $\mathbf{\ov 5^H}$, one needs the presence of at least one, but potentially more $U(1)$ stacks. The $\mathbf{\ov 5}$, $\mathbf{5^H}$ and $\mathbf{\ov 5^H}$ appear then as bi-fundamentals at intersections of the $U(5)$ stack and one of the $U(1)$ stacks. The singlets under the $SU(5)$ arise at intersections between different $U(1)$-stacks.

The perturbative superpotential is given by all gauge invariant couplings that are uncharged under all the global $U(1)$'s, the remnants of the Green-Schwarz mechanism. While the couplings 
\begin{align}
\mathbf{10} \mathbf{\ov 5} \mathbf{\ov 5^H}  \qquad  \qquad \mathbf{\ov 5} \mathbf{ 5^H} \mathbf{ 1}  \qquad \qquad    \mathbf{\ov 5^H} \mathbf{ 5^H} 
\end{align}
can in principle be perturbatively realized the other desired coupling
\begin{align}
\mathbf{10}_{(2,0)} \mathbf{10}_{(2,0)} \mathbf{5^H}_{(1,1)}
\end{align}
is perturbatively forbidden since it violates the global selection rules. Here the subscripts denote the charge of the respective matter fields under the global $U(1)$'s namely the one originating from the $U(5)_a$ and the $U(1)$ under which the Higgs field $\mathbf{5^H}$ is charged. Obviously, this coupling is not neutral under these two global $U(1)$'s and therefore perturbatively forbidden.

Recently, it has been realized that D-instantons carry charge under these global $U(1)$'s \cite{Blumenhagen:2006xt,Ibanez:2006da,Florea:2006si} (for recent reviews, see \cite{Blumenhagen:2009qh,Bianchi:2009ij}). For a specific product of matter fields they can compensate for the overshoot in the global $U(1)$ charge and induce the perturbatively missing couplings. For a rigid $O(1)$ instanton, which satisfies the severe constraints on the uncharged zero mode structure \cite{Argurio:2007qk,Argurio:2007vq,Bianchi:2007wy,Ibanez:2007rs}, the charge under the global $U(1)_x$ arising from a stack of $N_x$ branes wrapping the cycle $\pi_X$ in the internal manifold is given by\footnote{Other instanton configurations, such as multi-instantons \cite{Blumenhagen:2007bn,GarciaEtxebarria:2007zv,Cvetic:2008ws,GarciaEtxebarria:2008pi}  and so called rigid $U(1)$ instantons \cite{Aganagic:2007py,GarciaEtxebarria:2007zv,Petersson:2007sc,GarciaEtxebarria:2008iw,Ferretti:2009tz,Cvetic:2009ez} can induce superpotential terms.}
\begin{align}
Q_x = - N_x\,\, \pi_{E} \circ \pi_{x}\,\,.
 \end{align}
Here $\pi_E$ denotes the orientifold invariant cycle wrapped by the D-instanton.  The nonperturbative generation of Yukawa couplings via a rigid $O(1)$ instanton has been explicitly discussed in \cite{Blumenhagen:2007zk}.

In this work we extend the analysis of supersymmetric $SU(5)$ GUT's in a global context performed in \cite{Kiritsis:2009sf}, in which the authors pointed out that the most economical realization based on two stacks of D-branes poses severe phenomenological drawbacks. We investigate whether these drawbacks can be resolved by allowing additional D-brane stacks and find that the Georgi-Glashow model can be accommodated via three D-brane stacks, overcoming all problems encountered in the two D-brane stack realization. 
We present a global realization based on rational conformal field theory (RCFT), that admits a rigid $O(1)$ instanton inducing the perturbatively missing coupling $\mathbf{10} \mathbf{10} \mathbf{5^H}$. In contrast to Georgi-Glashow  D-brane realizations, flipped $SU(5)$ D-brane realizations, even with multiple $U(1)$ stacks have severe phenomenological problems. Specifically, the intriguing flipped SU(5) breaking mechanism of the GUT gauge symmetry down to the standard model gauge symmetry cannot lead to a consistent low energy theory without requiring the presence of additional geometric symmetries.

The paper is organized as follows. In section \ref{chap SU5} we analyze the Georgi-Glashow realization via $2$ and $3$ D-brane stacks in a bottom-up fashion. At the end we present a global three stack realization based on RCFT that exhibits a rigid $O(1)$ instanton that induces the desired coupling  $\mathbf{10} \mathbf{10} \mathbf{5^H}$. In section \ref{chap flipped SU5} we perform an analogous analysis for the flipped $SU(5)$ model. We conclude with some final remarks in section \ref{chap  conclusion}. In the appendices \ref{app string consistency conditions} and \ref{app GG 3 stacks} we lay out the basic ingredients for our systematic bottom-up analysis and present the results for the three-stack Georgi-Glashow realization. In appendix \ref{app global model} we present all details of the global realizations including the spectrum of the hidden sector.

\section{Georgi-Glashow model
\label{chap SU5}}
Before turning to D-brane realizations of the Georgi-Glashow model let us briefly introduce the usual supersymmetric SU(5)-GUT model. Later on we discuss potential D-brane realizations of it. The embedding of the standard model fields is displayed in Table \ref{table  SU(5) model}.

\begin{table}[h] \centering
\begin{tabular}{|c|c|c|c|}
\hline
 Representation &  SM matter embedding  & Multiplicity\\
\hline \hline
 $\mathbf {10}$                            &  $(q_L, u_R, e_R)$ & $3$  \\
\hline
 $\mathbf{\ov 5}$                              & $( L, d_R)$  & $3$  \\
\hline
 $ \mathbf{1}$                            & $\nu_R$  & $3$ \\
\hline
$\mathbf{5_H }+ \mathbf{  \ov 5_H}$  & $(H_u, T_u) + (H_d, T_d) $  & $1 + 1$  \\
\hline
\end{tabular}
\caption{\small {Spectrum for the supersymmetric $SU(5)$ model.}} 
\label{table  SU(5) model}
\end{table}\vspace{5pt}
The superpotential is given by
\begin{align}
W=\mathbf {10}\, \mathbf{  \ov 5}  \, \mathbf{  \ov 5^H} +  \mathbf {10}\,  \mathbf {10}\,  \mathbf {5^H} +  \mathbf{ \ov 5}  \, \mathbf{   5^H} \mathbf{1}  +
\mathbf{   5^H} \, \mathbf{ \ov 5^H}\,\,,
\end{align}
where the Yukawa coupling $\mathbf {10}\, \mathbf{  \ov 5}  \, \mathbf{  \ov 5^H}$  induces
the down-flavour quark masses, the coupling  $\mathbf {10}\,  \mathbf {10}\,  \mathbf {5^H}$ the up-flavour quark and charged lepton masses, respectively, and the coupling $ \mathbf{ \ov 5}  \, \mathbf{  5^H} \mathbf{1}$ the Dirac neutrino masses.

The breaking down to the standard model gauge groups occurs via an adjoint $\mathbf{24}$, which
acquires a vev, of the form
\begin{align}
\langle\, \mathbf{24}\,\rangle=\text{ diag} \left(v,v,v, -\frac{3}{2} v, -\frac{3}{2} v\right) \,\,,
\label{eq vev adjoint}
\end{align}
where $v$ is of the order $10^{16} \, GeV$. The hypercharge $U(1)_Y$ is embedded in the $SU(5)$ and given by
\begin{align}  
U(1)_Y= \text{diag}\left(-\frac{1}{3}, -\frac{1}{3}, -\frac{1}{3}, \frac{1}{2}, \frac{1}{2}
\right)\,
\end{align}
which remains unbroken once the adjoint $\mathbf{24}$ acquires a vev of the type \eqref{eq vev adjoint}. After this brief introduction of the Georgi-Glashow model we turn to the D-brane realization of it.

\subsection{D-brane realization}
The most economical way to embed  the Georgi-Glashow $SU(5)$ GUT  in a D-brane configuration is via two stacks of D-branes $a$, $b$. Stack $a$ contains 5 D-branes while stack $b$ is just a single D-brane, giving rise to the  gauge symmetry $U(5)_a \times U(1)_b$. The abelian $U(1)_a$ and $U(1)_b$ are generically anomalous and become massive via the Green-Schwarz mechanism. Thus the resulting gauge symmetry is the desired $SU(5)$. The massive $U(1)_a$ and $U(1)_b$ survive as global symmetries in the low energy effective theory  and  have to be preserved by all  perturbative couplings.

In Table \ref{table Spectrum for SU5 model} we display the origin of the respective matter fields for the realization of the Georgi-Glashow $SU(5)$ model based on two stacks of D-branes. This chiral spectrum satisfies the string consistency conditions laid out in appendix \ref{app string consistency conditions}. Note that the hypercharge $U(1)_Y$ is a subgroup of $U(5)$ and thus is guaranteed to remain massless. Therefore tadpole cancellation is the only constraint one has to ensure. However, for chiral matter with the transformation property  displayed in Table \ref{table Spectrum for SU5 model} there is potentially a massless combination, satisfying the constraints \eqref{eq massless constraint non-abelian} and \eqref{eq massless constraint abelian}, given by
\begin{align}
U(1)_X = \frac{1}{4} U(1)_a -\frac{5}{4} U(1)_b\,\,.
\label{eq linear combiantion}
\end{align}
Let us stress that the conditions on the transformation properties of the chiral matter fields arising from tadpole cancellation and masslessness of a $U(1)$ derived in the Appendix \ref{app string consistency conditions} are just necessary constraints. Whether tadpoles are really cancelled and whether an abelian symmetry remains massless or not, depends crucially on the concrete global realization. 
\begin{table}[h] \centering
\begin{tabular}{|c|c|c|c|c|}
\hline
 Sector & Matter All &  Transformation & Multiplicity \\
\hline \hline
 $aa'$                            & $\mathbf{10}$  & $\Yasymm_a$ & $3$\\
\hline
 $ab$                            & $\mathbf{ \ov 5}$  & $(\overline{a},b)$  & $3 $  \\
\hline
 $ab'$                            & $\mathbf{5^H }+ \mathbf{  \ov 5^H}$  & $(a, b)  +  (\ov a, \ov b) $ & $1+ 1$  \\
\hline
$bb'$ & $\mathbf{1}$ & $\ov {\Ysymm}_b$ & $3$ \\
 \hline
\end{tabular}
\caption{\small {Chiral spectrum for a D-brane realization of the  SU(5) model.}} 
\label{table Spectrum for SU5 model}
\end{table}\vspace{5pt}

The perturbative realized Yukawa couplings are
\begin{align}
\mathbf{10}_{(2,0)} \, \mathbf{\ov 5}_{(-1,1)} \,  \mathbf{\ov 5^H}_{(-1,-1)} \qquad  \mathbf{\ov 5}_{(-1,1)}  \,\mathbf{1}_{(0,-2)}\,  \mathbf{ 5^H}_{(1,1)} \qquad  \mathbf{ 5^H}_{(1,1)} \, \mathbf{\ov 5^H}_{(-1,-1)}\,\,,
\end{align}
where the indices indicate the charge under the global $U(1)_a$ and $U(1)_b$ symmetries, respectively. The Yukawa  coupling
\begin{align}
\mathbf{10}_{(2,0)}\, \mathbf{10}_{(2,0)}\,\mathbf{5^H}_{(1,1)}\,\,
\end{align}
which contains the up-flavour quark coupling is perturbatively forbidden. An instanton with global $U(1)$ charge $(-5,-1)$ under $U(1)_a$ and $U(1)_b$ can induce the missing coupling. As shown in \cite{Blumenhagen:2007zk} one needs three different instantons with global $U(1)$ charge $(-5,-1)$ to give masses to all three families. Note though that the non-perturbative generation of the Yukawa coupling $\mathbf{10}\, \mathbf{10}\, \mathbf{5^H}$ suggests that the bottom quark is heavier than the top quark, which is in contrast to experimental observations.

The perturbative realization of the Dirac neutrino  mass term suggests that the neutrino masses are of the same order as the other matter field masses. However, experiments show that the neutrinos masses are $10^{-10}$ to $10^{-16}$ times smaller than the other matter field masses. The see-saw mechanism gives a natural explanation for the smallness of the neutrino masses. A necessary ingredient for the seesaw mechanism is a large Majorana mass term
\begin{align}
\mathbf{1}_{(0,-2)}\, \mathbf{1}_{(0,-2)}
\end{align}
which can be induced non-perturbatively by an instanton \cite{Blumenhagen:2006xt,Ibanez:2006da,Cvetic:2007ku,
Ibanez:2007rs, Antusch:2007jd,Cvetic:2007qj,Ibanez:2008my} with global $U(1)$ charges $(0,4)$. If the Majorana mass term is in the range $(10^{12}-10^{15} )\,GeV $ one obtains neutrino masses of the observed order $(10^{-2} - 1)\, eV$.

Let us comment on potential phenomenological drawbacks of this 2-stack quiver.
\begin{itemize}
\item[(1)] The perturbatively realized coupling  $\mathbf{10}\, \mathbf{\ov 5} \,  \mathbf{\ov 5^H} $ contains the Yukawa coupling giving masses to the down-flavour quarks. On the other hand the coupling $\mathbf{10}\, \mathbf{10}\, \mathbf{5^H}$, which contains the Yukawa coupling that gives masses to the up-flavour quarks is perturbatively forbidden. It is induced by an instanton and thus suppressed compared to the coupling $\mathbf{10}\, \mathbf{\ov 5} \,  \mathbf{\ov 5^H} $. This suggests that for this quiver the bottom quark is heavier than the top quark which is in contrast to experimental observations.
\item[(2)] The instanton inducing the perturbatively missing Yukawa coupling $\mathbf{10}\, \mathbf{10}\, \mathbf{5^H}$ also generates the dangerous dimension $5$ operator $\mathbf{10}\, \mathbf{10}\, \mathbf{10}\, \mathbf{\ov 5}$ \cite{Kiritsis:2009sf}. For the Georgi-Glashow $SU(5)$ model the  dimension $5$ operator $\mathbf{10}\, \mathbf{10}\, \mathbf{10}\, \mathbf{\ov 5}$ includes $q_L\, q_L\, q_L\, L$  and $u_R\, u_R\, d_R\, E_R$, which if not sufficiently suppressed lead to a disastrous proton decay rate. Since the Yukawa coupling $\mathbf{10}\, \mathbf{10}\, \mathbf{5^H}$ is responsible for the up-flavour quark coupling we expect only a minor suppression from the instanton, which is not enough to saturate the bounds on the proton lifetime.
\item[(3)] The chiral spectrum displayed in table \ref{table Spectrum for SU5 model} allows for a massless $U(1)_X$ given in equation
\eqref{eq linear combiantion}. A massless $U(1)_X$ directly contradicts observations, and furthermore its charge explicitly forbids
       Majorana neutrino masses, both perturbatively and non-perturbatively. This model may be viable
       if the $U(1)_X$ photon acquires a sufficiently large mass, and the mechanism responsible for that might
       also generate neutrino masses, but it would be preferable to achieve all that directly in string theory.
       This is possible if the $U(1)_X$ acquires a mass from axion mixing. Note that this can happen even though the conditions \eqref{eq massless constraint non-abelian} and \eqref{eq massless constraint abelian} are satisfied, since the latter are just necessary, but not sufficient, conditions to have a massless $U(1)_X$\footnote{As discussed in appendix \ref{app string consistency conditions} the condition of having a massless $U(1)$ imposes constraints on the cycles the D-branes wrap. However, these constraints imply restrictions on the transformation behaviour of the chiral matter. Nevertheless the latter are just necessary conditions not sufficient.}.

\end{itemize}
As we will show momentarily all these problems can be overcome if one allows for an additional $U(1)$ stack $c$. The $U(1)_c$ becomes again massive via the Green-Schwarz mechanism and survives only as a global symmetry. The second problem, namely that the instanton that induces the desired Yukawa coupling $\mathbf{10}\, \mathbf{10}\, \mathbf{5^H}$ generically also generates dimension five proton decay operators, can be avoided if the matter fields $\mathbf{\ov 5}$ carry some global charge $U(1)_b$ while the Higgs field $ \mathbf{5^H}$ is rather  charged under the global $U(1)_c$. As we show in appendix \ref{app GG 3 stacks} if we furthermore want to avoid the presence of R-parity violating couplings $\mathbf{10} \, \mathbf{ \ov 5} \, \mathbf{\ov 5}$ and $\mathbf{ \ov 5} \, \mathbf{5^H}$ there are only two different types of choices for the origins of the fields transforming non-trivially under the $SU(5)$. Within each choice there are different realizations of the Georgi-Glashow model, depending on the transformation behaviour of the right-handed neutrinos which are singlets under the $SU(5)$.
In appendix \ref{app GG 3 stacks} we derive all three-stack quivers that mimic the Georgi-Glashow model, pass all string consistency conditions as well as some minimal phenomenological requirements. 

In the following we will discuss for each type one representative. We first discuss the configuration in which the $\mu$-term is realized perturbatively

\subsubsection{Three-stack quiver with perturbatively realized $\mu$-term
\label{sec three-stack with mu-term}}

In table \ref{table spectrum SU5 model setup 1}  we display the origin of the respective matter fields for a realization of the Georgi-Glashow $SU(5)$ model based on three stacks of D-branes. It corresponds to configuration $6$ in table \ref{table potential SU5 solutions pert mu} in appendix \ref{app GG 3 stacks}.
\begin{table}[h] \centering
\begin{tabular}{|c|c|c|c|c|}
\hline
 Sector & Matter All &  Transformation & Multiplicity\\
\hline \hline
 $aa'$                            & $\mathbf{10}$  & $\Yasymm_a$ & $3$ \\
\hline
 $ab$                            & $\mathbf{ \ov 5}$  & $(\overline{a},b)$  & $3 $  \\
\hline
 $ac$                            & $\mathbf{5^H }+ \mathbf{  \ov 5^H}$  & $(a, \ov c)  +  (\ov a, c) $ & $1+ 1$\\
\hline
$bc$ & $\mathbf{1}$ & $(\ov b, c)$ & $3$\\
 \hline
\end{tabular}
\caption{\small {Spectrum for $SU(5)$ three stack quiver with  perturbative $\mu$-term. }} 
\label{table spectrum SU5 model setup 1}
\end{table}\vspace{5pt}

In contrast to the realization based on two stacks of D-branes here we have only two perturbatively realized couplings, namely
\begin{align}
\mathbf{\ov 5}_{(-1,1,0)}  \,\mathbf{1}_{(0,-1,1)}\,  \mathbf{ 5^H}_{(1,0,-1)} \qquad  \mathbf{ 5^H}_{(1,0,-1)} \, \mathbf{\ov 5^H}_{(-1,0,1)}\,\,,
\end{align}
where the subscripts again denote the respective global $U(1)$ charges. The couplings
\begin{align}
\mathbf{10}_{(2,0,0)}\, \mathbf{10}_{(2,0,0)}\,\mathbf{5^H}_{(1,0,-1)}\,\,  \qquad \text{and} \qquad \mathbf{10}_{(2,0,0)} \, \mathbf{\ov 5}_{(-1,1,0)} \,  \mathbf{\ov 5^H}_{(-1,0,1)}
\end{align}
can be induced non-perturbatively via the instantons $E_1$ and $E_2$ which have global $U(1)$-charge $(-5,0,1)$ and $(0,-1,-1)$, respectively.
To get the desired hierarchy between the up-flavour and down-flavour quark masses we expect the ratio to be\footnote{
To be precise the instanton $E_1$ induces only masses for one up-flavour quark family, thus one needs two additional instantons giving masses to the other two up-flavour quark families. Generically they have different suppression factor. Thus the ratio $e^{-S^{E_1}_{ins}} : e^{-S^{E_2}_{ins}} \simeq 100$ explains the hierarchy between the top and bottom-quark mass.}
\begin{align}
e^{-S^{E_1}_{ins}} : e^{-S^{E_2}_{ins}} \simeq 100\,\,.
\end{align}
Note that the neutrino Dirac mass term is realized perturbatively. Thus in order to obtain the observed small neutrino masses we expect the presence of a large Majorana mass term for the right-handed neutrinos, which can be induced by an instanton $E_3$ carrying global $U(1)$ charge $(0,2,-2)$. With the string scale of the order $10^{18}\, GeV$
and a suppression factor  $e^{-S^{E_3}_{ins}}\simeq 10^{-5}$ we obtain via the seesaw mechanism neutrino masses in the observed range.

Let us now discuss if this setup indeed overcomes all the issues encountered for the 2-stack realization. First note that for the three-stack setup displayed in table \ref{table spectrum SU5 model setup 1} both couplings, the   $\mathbf{10}\, \mathbf{10}\, \mathbf{5^H}$ as well as $\mathbf{10}\, \mathbf{\ov 5} \,  \mathbf{\ov 5^H} $ are perturbatively forbidden. In case the suppression factor of the instanton inducing the latter one is larger than the suppression factor of the instanton generating the coupling $\mathbf{10}\, \mathbf{10}\, \mathbf{5^H}$ one gets the desired hierarchy between top and bottom quark masses. Furthermore, the instanton inducing the coupling $\mathbf{10}\, \mathbf{10}\, \mathbf{5^H}$ does not carry the right global charge to induce the dangerous dimension $5$ operator, which could lead to a disastrous proton decay rate. Finally, this setup does not satisfy the necessary conditions on having a massless $U(1)$, thus all linear combinations of $U(1)_a$, $U(1)_b$ and $U(1)_c$ are massive. That allows the presence of the Majorana mass term for the right-handed neutrino induced by a D-instanton, which was potentially forbidden in the 2-stack realization.

\subsubsection{Three-stack quiver with non-perturbative $\mu$-term
\label{sec three-stack no mu term}}
Again we discuss here only one representative of all the possible solutions displayed in appendix \ref{app GG 3 stacks}. The chiral spectrum of this quiver is displayed in table \ref{table spectrum SU5 model 3 stack setup 2} and corresponds to configuration $9$ in table \ref{table potential SU5 solutions non-pert mu} in appendix \ref{app GG 3 stacks}.
\begin{table}[h] \centering
\begin{tabular}{|c|c|c|c|c|}
\hline
 Sector & Matter All &  Transformation & Multiplicity\\
\hline \hline
 $aa'$                            & $\mathbf{10}$  & $\Yasymm_a$ & $3$ \\
\hline
 $ab$                            & $\mathbf{ \ov 5}$  & $(\overline{a},b)$  & $3 $  \\
\hline
 $ab'$                            & $\mathbf{ \ov 5^H}$  & $(\overline{a},\ov b)$  & $1 $  \\
\hline
 $ac'$                            & $\mathbf{5^H }$  & $(a,c)  $ & $1$\\
\hline
$bb'$ & $\mathbf{1}$ & $\Ysymm_b$ & $1$\\
 \hline
 $cc'$ & $\mathbf{1}$ &$\Ysymm_c$ & $2$\\
 \hline
\end{tabular}
\caption{\small {Chiral spectrum for SU(5) model based on three stacks of D-branes with non-perturbative $\mu$-term. }} 
\label{table spectrum SU5 model 3 stack setup 2}
\end{table}\vspace{5pt}

For this quiver the only perturbatively realized Yukawa coupling is
\begin{align}
\mathbf{10}_{(2,0,0)} \,\mathbf{\ov 5}_{(-1,1,0)} \, \mathbf{\ov 5^H}_{(-1,-1,0)}  \,\,,
\end{align}
which gives masses to the down-flavour quarks. The other desired couplings 
\begin{align} \nonumber
& \mathbf{10}_{(2,0,0)}\, \mathbf{10}_{(2,0,0)}\,\mathbf{5^H}_{(1,0,1)}  \qquad   \mathbf{ 5^H}_{(1,0,1)} \,  \mathbf{\ov 5^H}_{(-1,-1,0)} \\
\label{eq nonpert. couplings}
\\ \nonumber   & \mathbf{ \ov 5}_{(-1,1,0)} \,  \mathbf{5^H}_{(1,0,1)} \mathbf{ 1}_{(0,2,0)} \qquad   \mathbf{ \ov 5}_{(-1,1,0)} \,  \mathbf{ 5^H}_{(1,0,1)} \mathbf{1}_{(0,0,2)}
\end{align}
are induced via the D-instantons $E_1$, $E_2$, $E_3$ and $E_4$ which carry global $U(1)$ charges  
\begin{align}
E_1=(-5,0,-1)  \hspace{8mm} E_2 =(0,1,-1)  \hspace{8mm} E_3 = (0,-3,-1) \hspace{8mm} E_4 = (0,-1,-3)\,.
\end{align}
The first term in \eqref{eq nonpert. couplings}, induced by the instanton $E_1$, gives masses to the up-flavour quarks as well as to the charged leptons. Since the top-quark is the heaviest Standard model field particle the suppression of the instanton must be very small. The suppression factor of $E_2$ on the other hand should be rather large to account for a $\mu$-term of the order $100 GeV$. The instantons $E_3$ and $E_4$  induce the Dirac neutrino masses. Together with D-instantons that generate Majorana masses for the right-handed neutrinos they give via the see-saw mechanism the observed small neutrino masses\footnote{Large suppression factors of  $E_3$ and $E_4$ can  account for  the smallness of the neutrino mass \cite{Cvetic:2008hi}. In that case no Majorana masses for the right-handed neutrinos should be generated.}.

Let us again discuss whether this quiver indeed overcomes all the issues encountered for the two stack quiver. The major drawback of the two-stack quiver, namely that the instanton that induces the desired Yukawa coupling $\mathbf{10}\, \mathbf{10}\, \mathbf{5^H}$
also generates the dangerous dimension five operator $\mathbf{10}\, \mathbf{10}\,  \mathbf{10}\, \mathbf{ \ov 5}$, is not a problem for this quiver. Also this quiver like the 3-stack quiver discussed before does not exhibit any abelian symmetry that remains massless. Thus all perturbatively missing terms can be generated via D-instantons. However the perturbative realization of down flavour quark masses compared to the non-perturbative up flavour quark mass suggests exactly the opposite mass hierarchy compared to the observed one. Let us point out though that both couplings are further suppressed via world-sheet instantons which after all can in principle suppress the down-flavour quark masses such that the top-quark is indeed as observed the  heaviest standard model particle. The latter is however not the generic case and usually requires some amount of fine-tuning.

\subsection{Global realization of  a three-stack quiver
\label{sec global realization}}

Here we present a global realization of a three-stack quiver which is similar to the ones we discussed above. We will see that one can indeed find a rigid $O(1)$ instanton that exhibits the correct zero mode structure to induce the perturbatively forbidden Yukawa coupling
$\mathbf{10}\, \mathbf{10}\, \mathbf{5^H}$. The model is based on RCFT which are called Gepner orientifolds.

Gepner orientifolds are constructed by replacing the geometric notion of curled extra dimensions to form a compact manifold, by an algebraic procedure where the internal sector consists of tensor products of ${\cal N}=2$ minimal superconformal models with total central charge $c=9$ \cite{Angelantonj:1996mw,Blumenhagen:1998tj, Blumenhagen:2003su,Blumenhagen:2004cg,Blumenhagen:2004qu,Blumenhagen:2004hd} \footnote{For some initial studies on closed Gepner  constructions see \cite{Gepner:1987vz,Gepner:1987qi,Eguchi:1988vra}.}. In this context, there has been an extensive search for all possible embeddings of the standard model gauge theory in D-brane configurations \cite{Dijkstra:2004ym, Dijkstra:2004cc, Anastasopoulos:2006da, Kiritsis:2008ry}.

Before presenting a concrete example let us describe our search for realistic three-stack quivers, which is based on a previous study performed in\cite{Anastasopoulos:2006da}. There the authors searched first for local  configurations of D-brane boundary states that reproduce the chiral spectrum of the MSSM or extensions of it, such as $SU(5)$ GUT's. These local D-brane boundary state configurations were required to not saturate the tadpole constraints and moreover give rise to a hypercharge embedding that is compatible with the MSSM hypercharge assignment and does not become massive via the Green-Schwarz mechanism. In a few cases the D-brane boundary state configuration that gives rise to the MSSM, called visible sector in the following, was enough to satisfy the tadpole constraints. However, generically one needs additional boundary states to cancel the tadpoles. In the search performed in \cite{Anastasopoulos:2006da} the authors required that these additional boundary states, usually called hidden sector, are added in such a way that one does not have any chiral matter fields charged with respect to gauge groups in the visible and hidden sector, simultaneously. With this approach the authors found many globally consistent configurations, that give realistic chiral spectra, where the latter include also SU(5) GUT and  Pati-Salam-realizations. 
 
In this work we follow a similar path. We take the subset of local configurations that give rise to a $SU(5)$ GUT-like spectrum and analyze the superpotential by looking at the global $U(1)$ charges of the respective matter fields. In case a desired Yukawa coupling is missing we are looking for a rigid $O(1)$ instanton that has the correct zero mode structure to induce the missing coupling. To be more precise we identify all instanton boundary states that are orientifold invariant and do not exhibit any additional neutral fermionic zero modes apart from the two universal $\theta^{\alpha}$ modes \cite{Argurio:2007qk,Argurio:2007vq,Bianchi:2007wy,Ibanez:2007rs}. Then we further require that the intersection pattern of this instanton boundary state with the visible D-brane boundary state configuration is in such a way that it gives the correct charged zero mode structure to induce the perturbatively forbidden, but desired coupling.

Once such an instanton is found we are looking for a hidden sector that cancels the tadpoles in such a way that it does not intersect with the instanton. This way it is ensured that the instanton does not exhibit any additional zero modes charged with respect to the hidden sector, which would kill the instanton contribution to the perturbatively missing coupling. In general, an already known solution to the tadpole cancellation condition is not very likely to satisfy this criterion, so one usually has to perform a new search for hidden sectors, imposing the instanton zero-mode constraint. Note that this constraint is rather strong, since it demands complete absence of zero-modes, even vector-like ones.

In addition to intersecting the instanton brane, the hidden sector branes may also intersect the observable matter
brane. This is usually indeed what happens, although there are rare examples where there is no massless observable-hidden matter at all \cite{Anastasopoulos:2006da}, or where no hidden sector is needed to cancel all tadpoles \cite{Dijkstra:2004cc}. In the latter case, the problem is of course already solved, but then one looses the possibility of using the hidden sector for supersymmetry breaking. Apart from these rare cases, the common procedure is to allow observable-hidden matter, provided it is completely vector-like with respect to the entire
gauge group. In that case, the additional matter may acquire a mass without any gauge symmetry breaking. In principle, one may relax the above restriction further and even allow observable-hidden matter that is  chiral, but that becomes vector-like if the hidden sector gauge symmetry is removed. Then it depends on further details of the hidden sector dynamics what ultimately happens to the exotic observable-hidden matter, but it is not difficult to think of scenarios where it becomes sufficiently massive. In explicit examples, the number of tadpole solutions increases by several orders of magnitude under these relaxed conditions in comparison to  the strict ones ({\it i.e.} those allowing only vector-like observable-hidden matter). Our attitude here is that chiral observable-hidden matter of this kind is a lesser evil than superfluous instanton zero modes, and therefore we use the relaxed condition, after checking that the strict one does not generate any solutions. Previous searches have shown that under the strict  observable-hidden conditions, instantons with the correct zero mode structure to generate desired interactions \cite{Ibanez:2007rs,Kiritsis:2009sf} are very rare.

In \cite{Anastasopoulos:2006da} the authors found 7 different semi-realistic  three-stack realizations of the $SU(5)$ GUT's. In this search they allowed also for $O(1)$ gauge groups for the additional third D-brane stack. Let us also point out that this subset mainly contains setups in which the $\mu$ term is perturbatively forbidden \footnote{In the search performed in \cite{Anastasopoulos:2006da} three-stack quivers with a perturbatively realized $\mu$-term were considered as two stack quivers, rather than three stack quivers, due to the fact that the chiral spectrum arises from only two stacks. From the available data we can therefore not decide if a third brane with the right properties can be found.}. Thus configurations of the type discussed \ref{sec three-stack with mu-term} are not contained in the search we will perform. We leave it for future work to extend the search of three-stack quivers by also including quivers in which the $\mu$-term is perturbatively realized. 

Performing the analysis in the fashion described above we find one type of configuration that gives rise to a semi realistic model and 
exhibits a rigid $O(1)$ instanton generating the perturbatively forbidden Yukawa coupling $\mathbf{10}\, \mathbf{10}\, \mathbf{5^H}$.
The visible sector consists of three stacks of branes giving rise to the gauge symmetry
\begin{align}
U(5)_a \times U(1)_b \times O(1)_c\,\,.
\end{align}
The spectrum of a specific model is displayed in table \ref{table global model} , where any constituent D-brane boundary state of the hidden sector
is denoted by $h$. For the sake of clarity we do not display any specifics of the hidden sector. For the details, such as the hidden sector gauge symmetry as well as the spectrum within the hidden sector we refer to appendix \ref{app global model}.

\begin{table}[h] \centering
\begin{tabular}{|c|c|c|c|c|}
\hline
 Sector & Matter All &  Transformation & Multiplicity\\
\hline \hline
 $aa'$                            & $\mathbf{10}$  & $\Yasymm_a$ & $3$ \\
\hline
 $ab$                            & $\mathbf{ \ov 5}$  & $(\overline{a},b)$  & $3 $  \\
\hline
 $ab'$                            & $\mathbf{ \ov 5^H}$  & $(\overline{a},\ov b)$  & $3 $  \\
\hline
 $ac$                            & $\mathbf{5^H}$  & $(a,  c)  $ & $3$\\
 \hline
 \hline
 $a h$ & $\mathbf{5}^{\text{ex}}$ &  $(a, h)  $ &   4\\ \hline
 $a h$ &${ \mathbf{\ov 5}}^{\text{ex}}$ &  $(\ov a, h)  $ &   4\\\hline
 \end{tabular}
\caption{\small {Visible spectrum of a globally consistent SU(5) model based on three stacks of D-branes. }} 
\label{table global model}
\end{table}\vspace{5pt}

Note that for this particular configuration there are three pairs of Higgs and no neutrinos. Moreover the hidden sector intersects chirally with the visible sector, giving rise to exotics. The net number of $SU(5)$ exotics is as expected zero, thus they are non-chiral with respect to the GUT gauge symmetry. However, the exotics carry different charge with respect to the hidden gauge groups. They only acquire mass after a breakdown of the hidden gauge group or via D-instantons inducing mass terms for them. 

Let us turn to the Yukawa couplings. The only perturbatively realized coupling is
the down flavour coupling
\begin{align}
\mathbf{10}_{(2,0,0)} \,\mathbf{\ov 5}_{(-1,1,0)} \, \mathbf{\ov 5^H}_{(-1,-1,0)}.
\end{align} 
For the up flavour Yukawa coupling
\begin{align}
\mathbf{10}_{(2,0,0)}\, \mathbf{10}_{(2,0,0)}\,\mathbf{5^H}_{(1,0,1)}
\end{align}
which is perturbatively absent we find an instanton that wraps a rigid orientifold invariant cycle that has the following  intersection pattern with the visible branes 
\begin{align}
\pi_{E} \circ \pi_{a} = 1 \qquad  \pi_{E} \circ \pi_{b} = 0 \qquad \pi_{E} \circ \pi_{c} = 1 \,\,.
\end{align}
Thus it gives the correct uncharged and charged zero mode structure to induce the desired but perturbatively missing coupling. Note also that this instanton does not intersect with any of the hidden D-brane boundary states and thus does not exhibit any zero modes charged with respect to the hidden sector that would spoil the generation of $\mathbf{10} \mathbf{10}\mathbf{5^H}$. The suppression of the instanton turns out to be too large to account for the observed masses of the standard model. Let us stress though that this analysis is performed at the exact RCFT point and moving away from this exact point in moduli space might improve the situation.

Let us turn to the $\mu$-term 
\begin{align}
\mathbf{5^H}_{(1,0,1)} \, \mathbf{\ov 5^H}_{(-1,-1,0)}
\end{align}
which is perturbatively forbidden and can be generated by an instanton with the intersection pattern 
\begin{align}
\pi_{E} \circ \pi_{a} = 0 \qquad  \pi_{E} \circ \pi_{b} = -1 \qquad \pi_{E} \circ \pi_{c} = 1 
\end{align}

In order to be compatible with phenomenology one expects the instanton to exhibit a large suppression factor. 
Unfortunately, for this specific example  we do not find any rigid $O(1)$ instantons with such intersection pattern.

Despite its phenomenological problems, such as absence of $\mu$-terms and Dirac mass term for the neutrinos, a highly suppressed top-quark mass, as well as the presence of additional exotics that are chiral with respect to the hidden sector but not with respect to the SU(5), this configuration serves as global realization of the $SU(5)$ quivers discussed above in which one can find an instanton that satisfies the severe zero mode constraints to induce a perturbatively missing coupling $\mathbf{10} \mathbf{10}\mathbf{5^H}$. Let us stress again that we looked only in a small phenomenologically interesting subset of $SU(5)$ quivers while other appealing quivers with a perturbatively realized $\mu$-term were not covered by this search. Additionally, we only allowed for rigid $O(1)$ instantons to give non-perturbative contributions to the superpotential. However, as shown in \cite{Blumenhagen:2007bn,GarciaEtxebarria:2007zv,Cvetic:2008ws,GarciaEtxebarria:2008pi}  also multi-instanton configurations and so called rigid $U(1)$ instantons \cite{Aganagic:2007py,GarciaEtxebarria:2007zv,Petersson:2007sc,GarciaEtxebarria:2008iw,Ferretti:2009tz,Cvetic:2009ez} can generate some of the missing Yukawa couplings. We leave it for future work to extend the here performed analysis by extending the class of local configurations and by including additional effects for the non-perturbative generation of desired couplings.

\section{Flipped $SU(5)$ model
\label{chap flipped SU5}}
In this section we discuss the realization of supersymmetric flipped $SU(5)$-GUT models in orientifold models. Before we present and analyze specific D-brane configurations which give rise to flipped $SU(5)$ gauge theory in four-dimensional space-time let us give  a brief  introduction to the flipped $SU(5)$ model. It consists of a non-abelian part $SU(5)$ accompanied with an abelian $U(1)_X$ gauge symmetry. The standard model matter fields appear as antisymmetric $\mathbf {10_{\frac{1}{2}}}$, anti-fundamental $\mathbf{ \ov 5_{-\frac{3}{2}}}$ and singlet $\mathbf{ 1_{\frac{5}{2}}}$ under the $SU(5)$, where the subscript denote the charge of the respective representation under the $U(1)_X$.  In table \ref{table flipped SU(5) models} we present the embedding of the standard model fields into the flipped SU(5) multiplets. In addition to the electroweak Higgs fields $\mathbf{5_H }$ and $\mathbf{  \ov 5_H}$ the flipped $SU(5)$ model also contains the Higgs fields $\mathbf{10_H }$ and  $\mathbf{  \overline{10}_H}$, whose presence is crucial for the breaking mechanism of the GUT gauge symmetry down to the Standard model gauge symmetry.
\begin{table}[h] \centering
\begin{tabular}{|c|c|c|c|}
\hline
 Representation &  SM matter embedding  & Multiplicity & $U(1)_X$\\
\hline \hline
 $\mathbf {10}$                            &  $(q_L, d_R, \nu_R)$ & $3$ &$\frac{1}{2}$ \\
\hline
 $\mathbf{\ov 5}$                              & $( L, u_R)$  & $3$  & $-\frac{3}{2}$ \\
\hline
 $ \mathbf{1}$                            & $e_R$  & $3$& $\frac{5}{2}$  \\
\hline
$\mathbf{5_H }+ \mathbf{  \ov 5_H}$  & $(H_d, T_d) + (H_u, T_u) $  & $1 + 1$& $-1  \,\,\,\,\,\,\,\,1$  \\
\hline
$\mathbf{10_H }+ \mathbf{  \ov {10}_H}$  & $(\Delta) + (\ov \Delta)$  & $1 + 1$& $\frac{1}{2}  \,\,\,\,\, -\frac{1}{2}$  \\
\hline
\end{tabular}
\caption{\small {Spectrum for the D-brane realization of the flipped $SU(5)$ model.}} 
\label{table flipped SU(5) models}
\end{table}\vspace{5pt}
Note that the spectrum assignment is similar to the one of the Georgi-Glashow model with the exchange
\begin{align}
u_R \leftrightarrow d_R \qquad e_R \leftrightarrow \nu_r \qquad H_u \leftrightarrow H_d\,\,.
\end{align}
The hypercharge is a subgroup of $SU(5) \times U(1)_X$, given by
\begin{align}
U(1)_Y=-\frac{1}{5}\text{diag}\left(-\frac{1}{3}, -\frac{1}{3}, -\frac{1}{3}, \frac{1}{2}, \frac{1}{2}
\right) + \frac{2}{5} U(1)_X\,\,.
\end{align}
In addition to the gauge symmetries there is a discrete $\mathbf{Z}_2$ symmetry $  \mathbf {10^H} \rightarrow -  \mathbf {10^H}$. Then the superpotential takes the form
\begin{align}
W= \mathbf {10}\, \mathbf{  \ov 5}  \, \mathbf{  \ov 5^H} +  \mathbf {10}\,  \mathbf {10}\,  \mathbf {5^H} +  \mathbf{ \ov 5}  \, \mathbf{   5^H} \mathbf{1}+ \mathbf {5^H}  \, \mathbf{  \ov 5^H} + \mathbf {10^H}\,  \mathbf {10^H}\,  \mathbf {5^H} +
\mathbf {\ov {10}^H}\,  \mathbf {\ov {10}^H}\,  \mathbf {\ov 5^H}\,\,.
\end{align}
Here the first three terms give masses to the standard fields after the electroweak Higgses acquire a vev, the fourth term is the $\mu$-term, and the last two terms are crucial for the doublet-triplet splitting after the component $\Delta_{45}$ and $\ov \Delta_{45}$ of $\mathbf {10^H}$ and $\mathbf {\ov{10}^H}$, respectively, acquire a vev of the GUT scale.

\subsection{D-brane realization}
Again the most economical way to embed the flipped $SU(5)$ model in a
D-brane configuration is via two stacks of D-branes $a$, $b$. Stack
$a$ contains 5 D-branes while stack $b$ consists of just a single D-brane.
Thus the resulting gauge symmetry is then  $U(5)_a \times U(1)_b$,
where the abelian $U(1)_a$ and $U(1)_b$ are generically anomalous
and become massive via the Green-Schwarz mechanism. However, in 
order to mimic the flipped $SU(5)$ model the linear combination
\begin{align}
U(1)_X= \frac{1}{4} U(1)_a -\frac{5}{4} U(1)_b
\end{align}
has to remain massless, thus has to satisfy the  constraints \eqref{eq massless constraint non-abelian} and \eqref{eq massless constraint abelian} displayed in appendix \ref{app string consistency conditions}. In table \ref{table Spectrum for flipped SU(5) models} we display the origin of the respective matter fields for the realization of the flipped $SU(5)$ model based on two stacks of D-branes.
\begin{table}[h] \centering
\begin{tabular}{|c|c|c|c|c|}
\hline
 Sector & Matter All &  Transformation & Multiplicity & $U(1)_X$\\
\hline \hline
 $aa'$                            & $\mathbf{10}$  & $\Yasymm_a$ & $3$& $\frac{1}{2}$ \\
\hline
 $ab$                            & $\mathbf{ \ov 5}$  & $(\overline{a},b)$  & $3 $ & $-\frac{3}{2}$ \\
\hline
 $ab'$                            & $\mathbf{5^H }+ \mathbf{  \ov 5^H}$  & $(a, b)  +  (\ov a, \ov b) $ & $1+ 1$& $-1  \,\,\,\,\,\,\,\,\, 1$  \\
\hline
$bb'$ & $\mathbf{1}$ & $\ov {\Ysymm}_b$ & $3$& $\frac{5}{2}$  \\
 \hline
 $aa'$                            & $\mathbf{10^H} + \mathbf{\ov{10}^H}$  & $\Yasymm_a \,\,\,+\,\,\, \ov \Yasymm_a$ & $1+1$& $\frac{1}{2} \,\,\,\,\,-\frac{1}{2} $ \\
\hline
\end{tabular}
\caption{\small {Chiral spectrum for the flipped SU(5) model.   }} 
\label{table Spectrum for flipped SU(5) models}
\end{table}\vspace{5pt}

Let us again discuss the superpotential terms, beginning with the terms that give eventually masses to the standard model fields. The perturbatively realized couplings are
\begin{align}
\mathbf{10}_{(2,0)} \, \mathbf{\ov 5}_{(-1,1)} \,  
\mathbf{\ov 5^H}_{(-1,-1)} \qquad  \mathbf{\ov 5}_{(-1,1)}  \,\mathbf{1}_{(0,-2)}\, 
 \mathbf{ 5^H}_{(1,1)} \qquad  \mathbf{ 5^H}_{(1,1)} \, \mathbf{\ov 5^H}_{(-1,-1)}\,\,.
\end{align}
They contain the Yukawa couplings that give masses to the
up-flavour quarks, the charged leptons as well as the neutrinos and
also the $\mu$-term. Here the subscripts denote again the charges under the
global $U(1)$'s. However the coupling
\begin{align}
\mathbf{10}_{(2,0)} \, \mathbf{10}_{(2,0)} \, \mathbf{5^H}_{(1,1)}
\end{align}
whose presence is required to give  masses to the down-flavour quarks is
perturbatively forbidden. It can be generated by an instanton which
carries the  charge $(-5,-1)$ under the global $U(1)$'s. Since the
instanton induced Yukawa matrix factorizes one needs three different
instantons to generate masses for all three families. Thus the
non-perturbative nature of the coupling $\mathbf{10} \,
\mathbf{10}\, \mathbf{5^H} $ cannot only explain the observed
mass hierarchy between top and bottom quarks but also potentially 
explains the hierarchy between the different down-flavour families.

Let us now turn to the superpotential terms which are  crucial for
the GUT-breaking down to the Standard model gauge symmetry. Both
terms
\begin{align}
\mathbf{10^H}_{(2,0)} \, \mathbf{10^H}_{(2,0)} \,
\mathbf{5^H}_{(1,1)}  \qquad \qquad  \mathbf{\ov{10}^H}_{(-2,0)} \,
\mathbf{\ov{10}^H}_{(-2,0)} \, \mathbf{\ov 5^H}_{(-1,-1)}
\end{align}
whose presence is crucial are perturbatively forbidden. While the
first one will be generated by the same instanton which also
generates the Yukawa coupling $\mathbf{10} \, \mathbf{10}\,
\mathbf{5^H}$\footnote{This coupling will be actually induced by
a fourth instanton with the same charge under the global U(1)'s as the
instantons inducing the Yukawa coupling $\mathbf{10} \,
\mathbf{10}\, \mathbf{5^H}$. Note that the Higgs $\mathbf{10^H}$ are
basically a fourth family and thus in order to induce the Yukawa
coupling $\mathbf{10} \, \mathbf{10}\, \mathbf{5^H}$ for all three
families as well as for the Higgs fields $\mathbf{ 10^H}$ one needs
four different instantons with the global $U(1)$-charge $(-5,-1)$.
Note also that this potentially implies that one has to perform a
field redefinition in order to have the correct superpotential.
\label{footnote redefinition} } an instanton with charge
$(5,1)$ under the global $U(1)$ charges can induce the
perturbatively missing coupling $\mathbf{\ov{10}^H}\,
\mathbf{\ov{10}^H} \, \mathbf{\ov 5^H}$.

While this D-brane quiver after taking into account the
non-perturbative effects can in principle mimic the flipped $SU(5)$
model it has some phenomenological flaws, which we will discuss
below.
\begin{itemize}
\item[(1)] The perturbatively realized Yukawa coupling
$\mathbf{10} \, \mathbf{\ov 5} \,  \mathbf{\ov 5^H} $ contains the
Yukawa couplings giving masses to the up-flavour quarks and the
neutrinos. Thus they are expected to be of the same order which is
in contradiction to experiments that observe a hierarchy of
$10^{-16}$ between the top- quark mass and the neutrino masses. Note
though that this is not a problem due to the D-brane realization but
rather a problem within the flipped $SU(5)$ model. In
\cite{Antoniadis:1987dx} the authors present a flipped $SU(5)$ model
which allows for additional singlets $\Phi$, which are uncharged
under the $SU(5)$ as well as under the $U(1)_X$. These singlets
couple to the left-handed neutrinos via the coupling $\mathbf{ \ov {
10}^H}\, \mathbf{10} \, \Phi$. After the $\mathbf{ \ov { 10}^H}$
gets a vev on the order of the GUT-scale the Yukawa coupling
effectively becomes a large Majorana neutrino mass which via the
seesaw mechanism may explain the smallness of the neutrino masses.
However for the D-brane realization of the flipped $SU(5)$ model
with only two D-brane stacks one cannot accommodate a matter field
which is not charged under the $SU(5)$ and the $U(1)_X$\footnote{In principle the desired singlet could be an open string moduli transforming as an adjoint under the $U(1)_b$.}.

\item[(2)] As already discussed in \cite{Kiritsis:2009sf}
an instanton inducing the coupling $\mathbf{10} \, \mathbf{10}\,
\mathbf{5^H} $ also generates the dimension $5$ operator
$\mathbf{10} \, \mathbf{10}\, \mathbf{10}\, \mathbf{\ov 5}$. The
latter contains the dangerous dimension $5$ operator $q_L\, q_L\,
q_L\, L$  which if not sufficiently suppressed leads to a disastrous
proton decay rate. To match the observed hierarchy between the
top-quark and bottom-quark mass we expect the instanton suppression
on the order of $10^{-2}$, which is not enough to saturate the
bounds on the proton lifetime. Moreover, in the quiver displayed in
table \ref{table Spectrum for flipped SU(5) models} the dimension $5$
operator $\mathbf{10} \,  \mathbf{\ov 5}\,  \mathbf{\ov 5}\,
\mathbf{1} $ is perturbatively realized. This operator includes the
dimension five operator $u_R\, u_R\, d_R\, E_R$, which also has to
be highly suppressed to saturate the bounds on the proton lifetime.
Since it is perturbatively realized and thus only suppressed by the
string scale $M_s$ it poses a serious phenomenological problem and
predicts a proton lifetime not compatible with experimental
observations.

\item[(3)] The quiver displayed in table \ref{table Spectrum for flipped SU(5) models}
generically predicts the presence of the  terms
\begin{align}
\mathbf{10}_{(2,0)}\, \mathbf{\ov{10}^H}_{(-2,0)} \qquad \qquad
\mathbf{10^H}_{(2,0)}\, \mathbf{\ov{10}^H}_{(-2,0)}. \label{eq:
massterm}
\end{align}
Note that only one linear combination $\widetilde{\mathbf{10}} =
\sum_I c_I \mathbf{10}^I + c_H \, \mathbf{10^H}$, where $I$ runs
over all three families, becomes massive. However, independent on
whether the linear combination $\widetilde{\mathbf{10}}$ is
interpreted as the Higgs $\mathbf{10^H}$ or as one of the three
family matter fields $\mathbf{10}$ the presence of a mass term of
the form \eqref{eq: massterm} poses serious problems. In the latter
case it would induce a tadpole after $\mathbf{10^H}$ acquires a vev,
indicating an instability of the vacuum. For the former situation in
which $\widetilde{\mathbf{10}}$ is interpreted as the Higgs
$\mathbf{10^H}$ the mass term would forbid the simultaneous
acquirement of a vev for $\mathbf{10^H}$  and $\mathbf{\ov{10}^H}$,
otherwise supersymmetry is broken at the GUT scale.

Note, however that the mass term $ \mathbf{10^H} \, \mathbf{\ov{
10}^H}$ is induced via the three-point couplings
\begin{align}
\left(\Phi_{5} - \Phi_{1} \right) \, \mathbf{{\ov{10}}^H} \,
\mathbf{10^H}
\end{align}
where $\Phi_{5}$ and  $\Phi_{1}$ denote the scalar fields
transforming in the adjoint of the overall $U(1)$ of the $U(5)$
D-brane stack and of the $U(1)$ D-brane stack. These vevs are
related to the position in the internal space and in case they take
the same value the mass is zero and the problematic term \eqref{eq:
massterm} is absent. Generically it requires a large amount of
fine-tuning to avoid the presence of the mass terms of the form
\eqref{eq: massterm}.

\item[(4)] In the quiver displayed in table \ref{table Spectrum for flipped SU(5) models} the coupling
\begin{align}
\mathbf{10^H}_{(2,0)} \, \mathbf{\ov 5}_{(-1,1)} \,  \mathbf{\ov
5^H}_{(-1,-1)} \label{eq: dangerous Yukawa}
\end{align}
is perturbatively realized. Note that the field redefinition which
is necessary to ensure that only the Yukawa couplings $\mathbf{10}
\, \mathbf{10}\, \mathbf{5^H} $ and $\mathbf{10^H} \,
\mathbf{10^H}\, \mathbf{5^H} $ are present but no mixed terms
$\mathbf{10^H} \, \mathbf{10}\, \mathbf{5^H} $, which would lead to
large masses for some of the MSSM matter fields cannot ensure the
absence of the Yukawa coupling \eqref{eq: dangerous Yukawa} (see
footnote \ref{footnote redefinition}). However, after the component $\Delta_{45}$ of 
$\mathbf{10^H}$ acquires a vev of the GUT scale the presence of the
term \eqref{eq: dangerous Yukawa} would give rise to a large R-parity violating term $H_u \, L $, which is not compatible with experimental observations.
\end{itemize}
While the problem (3) can be avoided with some amount of fine-tuning the issues (1) and (2) can be overcome by allowing for another $U(1)$ brane stack, analogously to the Georgi-Glashow D-brane realization. However, even in a three-stack realization one faces the serious issue of the presence of the superpotential term $ \mathbf{10^H} \, \mathbf{\ov 5} \, \mathbf{\ov 5^H}$ that leads to the large R-parity violating term $L H_u$, thus giving $H_u$ and $L$ mass of the order of $M_{GUT}$, after the component $\Delta_{45}$ of $ \mathbf{10^H} $ acquires a vev of the GUT scale.

For  specific string compactifications there may exist additional
symmetries which emerge from the compactification
manifold. In case such a symmetry forbids the undesired couplings
the quiver displayed in table  \ref{table Spectrum for flipped SU(5) models} is a viable
D-brane configuration. However, let us emphasize that such symmetries may also
forbid some of the desired couplings. Moreover, for a generic
compactification we do not expect such symmetries to appear.

Summarizing we have shown that D-brane realization of the flipped
$SU(5)$ has serious phenomenological problems. Some of the problems 
can be overcome by allowing additional D-brane stacks. However for a generic string embedding 
the D-brane quivers mimicking the flipped $SU(5)$ model exhibit the
superpotential term  $ \mathbf{10^H} \, \mathbf{\ov 5} \,
\mathbf{\ov 5^H}$, that gives rise to an R-parity violating term of the order of $M_{GUT}$.

\section{Conclusions 
\label{chap  conclusion}}

In this work we discussed the realization of $SU(5)$ GUT's in the framework of orientifold compactifications. We analyze how in such compactifications the superpotential can be accommodated, where we assume that perturbatively non-realized couplings are generated via D-instanton effects. Often times the D-instanton that induces a desired Yukawa coupling also generates a coupling that poses phenomenological problems. For the $SU(5)$ orientifold realizations the coupling $\mathbf{10} \mathbf{10} \mathbf{5^H}$ is perturbatively forbidden, and thus needs to be realized non-perturbatively. However, in the most economical $SU(5)$ realization the D-instanton giving rise to the  $\mathbf{10} \mathbf{10} \mathbf{5^H}$ induces also the dangerous dimension $5$ operator $\mathbf{10} \mathbf{10} \mathbf{10} \mathbf{\ov 5}$. The presence of the latter would lead to a disastrous proton decay rate.

We show that this problem can be overcome by allowing for an additional D-brane stack. We display viable $SU(5)$ quivers based on three stacks of D-branes and investigate them with respect to their phenomenology. Furthermore, we present global Gepner model realizations of these quivers. These models exhibit  D-instantons that satisfy the severe constraints on the zero mode structure to induce the coupling $\mathbf{10} \mathbf{10} \mathbf{5^H}$. Unfortunately the instanton suppressions are too high to be phenomenologically viable. Nevertheless, these examples serve as global realization of the phenomenological viable $SU(5)$ quivers. The performed search of global realizations contained only a small subset of viable quivers and it would be interesting to extend the search by also allowing quivers with a perturbatively realized $\mu$-term.

Finally, we perform an analogous analysis for flipped $SU(5)$ models. In the absence of any additional geometric symmetries of the compactification manifold D-brane quivers mimicking the flipped $SU(5)$ model exhibit severe phenomenological problems. The  $\mathbf{10^H}$ required for the intriguing $SU(5)$ breaking mechanism generically couples to the standard model fields. After acquiring a vev of the GUT scale it induces large masses for the standard model fields not compatible with observations. 

\section*{Acknowledgements}
We would like to acknowledge interesting discussions and correspondence with M. Bianchi, M. Cveti{\v c}, J. Halverson, A. Lionetto, J.F. Morales, E. Kiritsis and N. D. Vlachos. P.A. would like to thank University of Crete, LP Ecole Normale Sup\'erieure de Lyon, Ecole Polytechnique and CERN for hospitality during the last stage of this work.  R.R. and A.N.S. are grateful to KITP Santa Barbara  for hospitality during parts of this work. This research is supported by the Dutch Foundation for Fundamental Research of Matter (FOM) as part of the program STQG (String Theory and Quantum Gravity, FP 57). This work has been partially supported by funding of the European Research and Training Network (RTN)
grant ``Unification in the LHC era ''(PITN-GA-2009-237920), the Spanish Ministerio de Ciencia e Innovaci\'on, Research Project FPA2008-02968, and by the Project CONSOLIDER-INGENIO 2010, Programme CPAN (CSD2007-00042). P.A was supported by FWF P22000 ``Aspects of String/Gauge Theory Duality''.


\appendix

\section{String consistency conditions and phenomenological constraints
\label{app string consistency conditions}}
In this appendix, we briefly summarize string consistency conditions that D-brane quivers have to satisfy. The latter contain constraints arising from tadpole cancellation and constraints that have to be fulfilled in case a linear combination $\sum_x  q_x U(1)_x$ should remain massless and thus survive as abelian gauge symmetry in the low energy effective action.  For a more detailed description, we refer the reader to \cite{Cvetic:2009yh} (for an analogous analysis see \cite{Bianchi:2000de,Dijkstra:2004cc})\footnote{For analogous work see \cite{Ibanez:2008my,Leontaris:2009ci,Anastasopoulos:2009mr,Kiritsis:2009sf,Cvetic:2009ez,Cvetic:2010mm,Anastasopoulos:2010ca, Cvetic:2010dz,Blumenhagen:2010dt,Fucito:2010dk}. First local (bottom-up) constructions were discussed in \cite{Antoniadis:2000ena,Aldazabal:2000sa,Antoniadis:2001np}.}. 

\subsection{Tadpole cancellation}

The tadpole cancellation condition, given by
\begin{align}
\sum_x N_x \left( \pi_x + \pi'_x\right) =4 \pi_{O}\,\,,
\label{eq tadpole}
\end{align}
is a condition on the cycles that the D-branes wrap. Here $\pi_x$, $\pi'_x$ and $\pi_{O}$ denote the homology class of the cycles the brane $x$, its orientifold image $x'$ and the orientifold $O$ wrap. Moreover, $N_x$ is the number of D-brane for stack $x$. Multiplying the tadpole cancellation condition with the homology class $\pi_a$ corresponding to the cycle wrapped by the D-brane stack $a$ and using the chiral spectrum displayed in table \ref{table chiral spectrum} one derives constraints on the transformation behaviour of the chiral matter given by
\begin{align}
\#(\fund_a) + (N_a-4)\#( \, \Yasymm_a) + (N_a+4) \#
(\Ysymm_a)=0 \,\,,\label{eq constraint1}
\end{align}
Note that for $N_a > 2$ this condition is the usual anomaly cancellation condition for non-abelian $SU(N_a)$ gauge symmetries. 
For $N_a=2$ it is a string-theoretic condition for anti-symmetric $U(2)$ tensors that does not correspond to any
field-theoretic anomaly condition. However, since these anti-symmetric tensors carry a charge under the phase
symmetry of $U(2)$, they can be distinguished from $SU(2)$ singlets. Therefore this
condition can be imposed on the field theory spectrum, and it {\it must} be imposed to have
any chance to find a string theory embedding. For $N_a=1$ the anti-symmetric tensor cannot even be
detected in the massless spectrum, and hence a given field theory spectrum may correspond to
a string theory spectrum with any number of chiral anti-symmetric tensors (where ``chiral" is defined 
as for $N_a > 2$), which are infinite towers with a vanishing ground state dimension.  However, since that number must be an integer, this still imposes a condition
\begin{align}
\#(\fund_a)+ 5 \# (\Ysymm_a)=0  \qquad \text{mod} \,3\,\,.
\label{eq constraint2}
\end{align}
    \begin{table}
\centering
\begin{tabular}{|c|c|}
\hline
Representation  & Multiplicity \\
\hline $ \Yasymm_a$
 & ${1\over 2}\left(\pi_a\circ \pi'_a+\pi_a \circ  \pi_{{\rm O}6} \right)$  \\
$\Ysymm_a$
     & ${1\over 2}\left(\pi_a\circ \pi'_a-\pi_a \circ  \pi_{{\rm O}6} \right)$   \\
$( \fund_a,{\overline \fund}_b)$
 & $\pi_a\circ \pi_{b}$   \\
 $(\fund_a, \fund_b)$
 & $\pi_a\circ \pi'_{b}$
\\
\hline
\end{tabular}
\vspace{2mm} \caption{Chiral spectrum}
\label{table chiral spectrum}
\end{table}

\subsection{Massless $U(1)$'s}
In order to have a massless linear combination $U(1)$\footnote{Note that higher-dimensional anomalies might affect the four-dimensional theory upon decompactifications and render masses to gauge bosons which are free of four dimensional anomalies \cite{Aldazabal:2000cn,Antoniadis:2002cs,Anastasopoulos:2003aj,Anastasopoulos:2004ga}.}
\begin{align}
U(1)=\sum_x \,q_x\, U(1)_x
\end{align}
the cycles that the D-brane stacks $x$ wrap have to satisfy \cite{Aldazabal:2000sa}
 \begin{align}
 \sum_{x} q_x N_x (\pi_x -\pi'_x) =0\,\,.
 \end{align}
Analogously to the tadpole cancellation, multiplying both sides with the homology class $\pi_a$ and using the relations displayed in table
\ref{table chiral spectrum} one obtains constraints on the transformation properties of the chiral matter.  
They take the form 
\begin{align}
 \sum_{x \neq a} q_x\,
N_x \#(\fund_a,{\ov \fund_x}) - \sum_{x \neq a} q_x\, N_x \#(\fund_a,\fund_x) = q_a\,N_a \,\Big(\#(\Ysymm_a) + \# (\, \Yasymm_a)\Big) \,\,
\label{eq massless constraint non-abelian}
\end{align}
for $N_a>1$. The case $N_a=1$ requires a little more care due to  the fact that in massless spectrum the antisymmetric tensor is absent. 
Using \eqref{eq constraint1} to express the ``would be" antisymmetrics in terms of the fundamentals and symmetrics one obtains
\begin{align}\sum_{x \neq a} q_x\, N_x \#(\fund_a,{\ov \fund_x}) -
\sum_{x \neq a} q_x\, N_x \#(\fund_a,\fund_x)= q_a\,\frac{\#(\fund_a) + 8 \#
(\Ysymm_a)}{3}  \,\,.
\label{eq massless constraint abelian}
\end{align}
Since  the flipped $SU(5)$ model requires an additional abelian gauge symmetry, namely $U(1)_X$ we require these constraints on the chiral spectrum of the flipped $SU(5)$ model to be satisfied by the linear combination $U(1)_X$.

\subsection{Derivation for RCFT models}

The foregoing derivations were made using the language of D-branes wrapping cycles on a manifold. Here
we will show how the same formulas can be derived using boundary and crosscap states on RCFT orientifolds.
For equation (\ref{eq constraint1}) it suffices to refer to \cite{Bianchi:2000de}. In \cite{Dijkstra:2004ym}
this was worked out for the simple current boundary state formalism developed in \cite{Fuchs:2000cm}.

Equation (\ref{eq massless constraint non-abelian}) can be derived as follows.
The condition that a massless $U(1)$ boson does not couple to an RR-axion is \cite{Dijkstra:2004ym}
\begin{align}
\sum_x q_x N_x(R_{x(m,J)} - R_{x^c(m,J)}) = 0\,\,,
\label{masslessu1}
\end{align}
where $q_x$ and $N_x$ are as above, and  $R_{x(m,J)}$ are the boundary coefficients as defined in \cite{Fuchs:2000cm}.
Here $x$ labels distinct boundary state, and $(m,J)$ labels Ishibashi states, closed string states that can propagate in the
transverse channel of an annulus, where $m$ refers to a state in the bulk theory, and $J$ is a degeneracy label. Which $m$'s  appear
and with which degeneracy is determined by the modular invariant partition function. Eqn.  (\ref{masslessu1}) must be satisfied
for every Ishibashi state which contains massless spinors\rlap.\footnote{In \cite{Dijkstra:2004ym} it is stated erroneously that this condition should
hold for {\it all} Ishibashi states. However,
in the actual standard model search presented in this paper, the condition was limited to Ishibashi states containing massless spinors.} This is determined only by $m$ and not by $J$.

In order to derive (\ref{eq massless constraint non-abelian}) it turns out that we need only a subset of these conditions, namely
\begin{align}
\sum_x q_x N_x (R_{x(m,J)} - R_{x^c(m,J)})  w_m = 0
\label{masslessu2}
\end{align}
where $w_m$  is the Witten index, counting the difference of  spinors and anti-spinors in a character.  No sum over $m$ is implied. This condition is a subset of (\ref{masslessu1}) because in general there are some Ishibashi states with an equal number of spinors and anti-spinors, which
would contribute to $U(1)$ mass, but not to (\ref{masslessu2}). Hence the condition we will derive is a necessary, but not sufficient
condition for a vanishing $U(1)$ mass.

We now perform a transformation from the transverse channel to the direct channel of the annulus, in a completely analogous
way as the derivation of cubic anomaly cancellation from tadpole cancellation.
The Witten indices transform exactly like characters, but are constants. Hence under this transformation we get $w_m=\sum_i w_iS_{im} $, where
$S$ is the modular transformation matrix.
Now we multiply the equations with a factor
\begin{align}
\sum_{J'} {R_{a(m,J')} g^{\Omega,m}_{J'J} \over S_{0m}}\,\,,
\label{masslessu3}
\end{align}
where $g^{\Omega,m}$ is the Ishibashi metric \cite{Fuchs:2000cm} on each degeneracy space. Finally we sum over $m$ and $J$ to obtain
\begin{align}
\sum_x q_x N_x \sum_i w_i \sum_{m,J',J} {S_{im} R_{a(m,J')} g^{\Omega,m}_{J'J} R_{x(m,J)}\over S_{0m}} - (x \rightarrow x^c) = 0\,\,.
\label{masslessu3}
\end{align}
The last sum is precisely the expression for the annulus coefficients, and hence we
get
\begin{align}
\sum_x q_x N_x \sum_i w_i (A^i_{~ax} - A^i_{~ax^c} ) = 0\,\,.
\label{masslessu3}
\end{align}
The contraction with the chiral characters $w_i$ turns this into the chiral intersection, {\it i.e} the first term is precisely  $\#(\fund_a,\fund_x)$
as defined  above.
This expression should hold for any boundary state label $a$. If one chooses a label $a$ that coincides with one of the labels $x$
which participates in the $U(1)$ symmetry of interests ({\it i.e.} $q_aN_a \not= 0$), then one may write
$$\sum_i w^i A^i_{~aa} = \frac12 \sum_i w_i (A^i_{~aa} + M^i_a) + \frac12 \sum_i w_i (A^i_{~aa} - M^i_a) = \Big(\#(\Ysymm_a) + \# (\, \Yasymm_a)\Big)$$
Note that the sum in (\ref{masslessu1}) is over {\it pairs} $(x,x^c)$ of conjugate boundary labels, labelled by $x$. Furthermore, if (\ref{masslessu3}) holds
for a label $a$ it automatically holds for its conjugate $a^c$, because $\sum_i w_i A^i_{~ax}=-\sum_i w_i A^i_{~a^cx^c}$. Hence we can use the same basis
of pairs $(a,a^c)$ for all boundary labels, and then only the case $x=a$ can occur.  The final result is then indeed precisely (\ref{eq massless constraint non-abelian}).

Note that using the completeness condition for boundaries \cite{Pradisi:1996yd}, $\sum_a R^*a_{m,J}  R^*a_{m',J'}  = \delta_{m,m'}\delta_{J,J'}$,  one can invert the derivation, so that
(\ref{masslessu2}) can be derived from (\ref{eq massless constraint non-abelian}). However (\ref{masslessu1}) does not follow, and hence, as already stated above, (\ref{eq massless constraint non-abelian})
 is in general only a necessary condition for masslessness of a $U(1)$. We have examined in a few cases how close it is to being sufficient. It turns out
that very often the quantity $R_{x(m,J)} - R_{x^c(m,J)}$ vanishes for all $x$ if the Witten index of $m$ is zero. This is true, for example, for
all modular invariant partition functions and all orientifolds of the tensor products $(3,3,3,3,3)$, $(3,8,8,8)$, $(6,6,6,6)$ and $(2,2,2,2,2,2)$. Hence in all these
cases (\ref{eq massless constraint non-abelian}) is actually sufficient, provided all boundary labels $a$ are taken into account. The tensor product $(4,4,10,10)$ provides
some examples where $R_{x(m,J)} - R_{x^c(m,J)}$ does not vanish if $w_m=0$, but only for a relatively small set of values of $x$. So cases where
(\ref{eq massless constraint non-abelian}) is not sufficient are rare.

The practical use of (\ref{eq massless constraint non-abelian}) is in determining if a postulated brane configuration has any chance of having a massless $U(1)$ boson (for example $Y$) in an explicit realization in string theory. Once one has found such a realization, one might as well check (\ref{masslessu1}) directly. Hence in practice the set of labels $a$ for which one uses it is just the set of branes appearing in the postulated brane configuration. Then certainly it is just a necessary, and not a sufficient condition. 

This condition also plays a r\^ole in the discussion of charge violation by instantons. Then $a$ is a candidate instanton brane, and the left-hand side of (\ref{masslessu3}) must be non-zero to get the required charge violation. Clearly a non-vanishing charge violation implies  that the corresponding $U(1)$ must be massive, but the converse is not necessarily true: for a massive $U(1)$ it may happen that there are no branes that violate  conservation of the charge. This was pointed out already in \cite{Ibanez:2006da} (in particular footnote 16 in that paper). We see now that this can only happen if there are contributions to the vector boson mass from Ishibashi states with a vanishing Witten index.

\subsection{Phenomenological requirements}
There are various phenomenological constraints which arise from experiments. We list them below.
\begin{itemize}
\item[$\bullet$]All the Yukawa couplings that give masses to the three families are realized, either perturbatively or non-perturbatively via D-instantons. Thus we require the presence of the terms   $\mathbf {10}\,  \mathbf {10}\,  \mathbf {5^H}$, $\mathbf {10}\, \mathbf{  \ov 5}  \, \mathbf{  \ov 5^H}$  and $ \mathbf{ \ov 5}  \, \mathbf{  5^H} \mathbf{1}$.
\item[$\bullet$]  For the flipped $SU(5)$ model we require the presence of terms $\mathbf {10^H}\,  \mathbf {10^H}\,  \mathbf {5^H}$ and 
$\mathbf {\ov {10}^H}\,  \mathbf {\ov {10}^H}\,  \mathbf {\ov 5^H}$ which are crucial for the breaking pattern of the flipped $SU(5)$ down to the standard model gauge symmetry.
\item[$\bullet$] We forbid any R-parity violating couplings $\mathbf {10}\, \mathbf{  \ov 5}  \, \mathbf{  \ov 5}$ or  $ \mathbf{ \ov 5}  \, \mathbf{  5^H}$ on perturbative or non-perturbative level. Specifically that implies that none of the instantons whose presence is required to induce some of the missing but desired couplings induces also the R-parity violating couplings.
\item[$\bullet$] We forbid the  presence of the dimension five operator $\mathbf {10}\, \mathbf {10}\,\mathbf {10}\,\mathbf{  \ov 5}$ on perturbative or non-perturbative level. For the  flipped $SU(5)$ model we also require the absence of the dimension five  operator $\mathbf {10}\,\mathbf{  \ov 5}\, \mathbf{  \ov 5}\, \mathbf{ 1}$, again on perturbative and non-perturbative level. As before that implies that none
of the instantons whose presence is required to induce some of the missing but desired couplings induces also these dimension five operators.
\item[$\bullet$] For the Georgi-Glashow D-brane realization often times an instanton which is required to generate a desired Yukawa coupling
also induces a tadpole $\mathbf{1}$ and thus an instability for the setup. We rule out any setup which requires the presence of such an instanton.
\end{itemize}

\section{Georgi-Glashow realizations based on three stacks
\label{app GG 3 stacks}}

In this appendix we present all 3-stack realizations of  the Georgi-Glashow model that satisfy all the string consistency conditions as well as all the phenomenological constraints laid out in appendix \ref{app string consistency conditions}. We distinguish between two different types of setups, for the first type  the $\mu$-term is perturbatively realized and for the second type the $\mu$-term is perturbatively forbidden and must be generated non-perturbatively.

In table \ref{table potential SU5 solutions pert mu} we display all possible solutions with exactly three right-handed neutrinos.
In the second line we display all possible origins for the matter fields\footnote{Note that this is true up to symmetries. For instance we take into account the symmetry under the exchange of stack $b$ with stack $c$. Moreover here we only display solutions with a perturbatively realized $\mu$-term.}. We find 12 different D-brane configurations, where solutions marked with a $*$ potentially exhibit a massless $U(1)$. 
In section \ref{sec three-stack with mu-term} we discuss in detail the phenomenology of the configuration 6.

\begin{table}[h]
\hspace{-.25cm}
\scalebox{.87}{
\begin{tabular}{|c|c|c|c|c|c|c|c|c|c|c|c|c|}\hline
        \multirow{2}{*}{Solution \#}&\multicolumn{1}{|c}{$\mathbf{10}$} & \multicolumn{1}{|c}{$\mathbf{\ov 5}$} & \multicolumn{1}{|c}{$\mathbf{ 5^H}$} & \multicolumn{1}{|c}{$\mathbf{\ov 5^H}$}  & \multicolumn{8}{|c|}{$\mathbf{1}$}
        \\ \cline{2-13}
        &$\Yasymm_a$&$(\ov{a},b)$&$(a,\ov c)$&$(\ov{a},c)$&$(b,\ov c)$ &$(\ov b, c)$ & $(b,c)$ & $(\ov b, \ov c)$ & $\Ysymm_b$ &  $\ov \Ysymm_b$ & $\Ysymm_c$ &  $\ov \Ysymm_c$
        \\
        \hline\hline
1&3&3&1&1&0&0&0&0&0&0&0&3\\ \hline
2&3&3&1&1&0&0&0&1&0&1&0&1\\ \hline
3&3&3&1&1&1&0&0&0&1&0&0&1\\ \hline
$4^*$&3&3&1&1&0&0&0&0&0&3&0&0\\ \hline
5&3&3&1&1&0&0&0&3&0&0&0&0\\ \hline
6&3&3&1&1&0&3&0&0&0&0&0&0\\ \hline
$7^*$&3&3&1&1&3&0&0&0&0&0&0&0\\ \hline
$8^*$&3&3&1&1&0&1&0&1&1&0&0&0\\ \hline
$9^*$&3&3&1&1&0&0&0&0&3&0&0&0\\ \hline
10&3&3&1&1&0&1&0&0&0&1&1&0\\ \hline
11&3&3&1&1&1&0&0&1&0&0&1&0\\ \hline
12&3&3&1&1&0&0&0&0&0&0&3&0\\ \hline
\end{tabular}}
\caption{\small 3-stack quiver realizations of the Georgi-Glashow model  with pert. $\mu$-term.
\label{table potential SU5 solutions pert mu}
}
\end{table}

In table \ref{table potential SU5 solutions non-pert mu} we display all possible 3 D-brane-stack realizations of the Georgi-Glashow model in which the $\mu$-term is not perturbatively realized. These satisfy the severe string consistency constraints as well as the phenomenological conditions laid out in the previous appendix. Again  solutions marked with a $*$ potentially exhibit a massless $U(1)$. In section \ref{sec three-stack no mu term} we discuss the configuration 9 as  a representative with respect to their phenomenology in  detail.

\begin{table}[h]
\hspace{-.15cm}
\scalebox{.87}{
\begin{tabular}{|c|c|c|c|c|c|c|c|c|c|c|c|c|}\hline
        \multirow{2}{*}{Solution \#}&\multicolumn{1}{|c}{$\mathbf{10}$} & \multicolumn{1}{|c}{$\mathbf{\ov 5}$} & \multicolumn{1}{|c}{$\mathbf{ \ov 5^H}$} & \multicolumn{1}{|c}{$\mathbf{5^H}$}  & \multicolumn{8}{|c|}{$\mathbf{1}$}
        \\ \cline{2-13}
        &$\Yasymm_a$&$(\ov{a},b)$&$(\ov a,\ov b)$&$(a,c)$&$(b,\ov c)$ &$(\ov b, c)$ & $(b,c)$ & $(\ov b, \ov c)$ & $\Ysymm_b$ &  $\ov \Ysymm_b$ & $\Ysymm_c$ &  $\ov \Ysymm_c$
        \\
        \hline\hline
1&3&3&1&1&0&0&0&1&0&0&0&2\\ \hline
$2^*$&3&3&1&1&0&0&0&0&0&2&0&1\\ \hline
3&3&3&1&1&1&0&1&0&0&0&0&1\\ \hline
$4^*$&3&3&1&1&0&0&0&2&0&1&0&0\\ \hline
5&3&3&1&1&1&0&0&1&1&0&0&0\\ \hline
6&3&3&1&1&0&0&1&0&2&0&0&0\\ \hline
7&3&3&1&1&1&0&0&0&0&1&1&0\\ \hline
8&3&3&1&1&0&0&2&0&0&0&1&0\\ \hline
9&3&3&1&1&0&0&0&0&1&0&2&0\\ \hline
\end{tabular}}
\caption{\small 3-stack quiver realizations of the Georgi-Glashow model  with non-pert. $\mu$-term.
\label{table potential SU5 solutions non-pert mu}
}
\end{table}

\section{Globally consistent 3-stack model
\label{app global model}}

Here we present globally consistent Gepner configurations that give rise to a Georgi-Glashow-like structure and exhibit an instanton that induces the perturbatively missing Yukawa coupling $\mathbf{10}\,\mathbf{10}\, \mathbf{5^H}$. We find two different types of global realizations and we present a representative of each here. Before doing so let us however explain the notation in the tables to come. 

In  the first column the table displays the whole number of states for a particular sector. The last column gives the net chirality of these states. The gauge groups are displayed in the first row, where a $V$ stands for fundamental, $\bar V$ for anti-fundamental, $S$ for the symmetric, $A$ for the anti-symmetric and $Ad$ for the adjoint under the respective gauge group. The column denoted by $Ins$ represents the rigid $O(1)$-instanton which will induce the  Yukawa coupling $\mathbf{10}\,\mathbf{10}\, \mathbf{5^H}$. Fields charged with respect to it denote the charged zero modes. One can easily see that the representatives below indeed exhibit the correct charged zero mode structure to induce the coupling  $\mathbf{10}\,\mathbf{10}\, \mathbf{5^H}$.

\subsection{Gepner Orientifold of Type I}

We find 6 global realizations with gauge group $ \mathbf{U( 5 )} \times \mathbf{ U( 1 )} \times  \mathbf{O( 1 )} \times O( 2 ) \times O( 1 ) \times O( 1 ) \times U( 1 ) \times O( 1 ) \times O( 1 ) \times U( 3 )$ where the first 3 gauge groups (the highlighted ones) denote the visible sector. The 6 different global realizations differentiate only in their massless spectrum in the hidden sector, the visible sector is for all 6 realizations the same. 

Let us specify the Gepner orientifold.  The internal sector of these models consists of a tensor product of four copies of ${\cal N}=2$ superconformal minimal models with levels $k_i=\{1,10,22,22\}$. This tensor product has 50 symmetric modular invariant partition functions. The one of our interest yields a closed string spectrum characterized by Hodge numbers $h_{11}=32$, $h_{12}=20$ and 237 singlets. These numbers identify it uniquely. This MIPF allows 4 different orientifold  choices, according to  the prescription given in \cite{Fuchs:2000cm}. The results below were obtained for one of these four (according to the labelling conventions used in \cite{Anastasopoulos:2006da} this case corresponds to MIPF nr. 26, orientifold nr. 1). A representative of these 6 realizations is displayed in table \ref{tab global model}.
 
Here we  divided the table into the following segments: the standard model fields (1-5), where the neutrinos arise from the non-chiral sector displayed in line 5, the instanton zero modes (6-7), chiral observable-hidden matter (8-15), non-chiral observable-hidden matter (16-23), non-chiral observable rank two tensors (24-30), chiral matter within the hidden sector (31-42), and non-chiral matter within the hidden sector (42-53).

Anti-symmetric tensors for $O(1)$ and $U(1)$ are shown even though 
their ground state dimension vanishes. The multiplicities of these sectors are however well-defined, and they
manifest themselves at higher excitation levels and, if they are chiral, through the tadpole cancellation condition \eqref{eq constraint1}. Note that anti-symmetric tensors {\it are} important if a matter brane is converted
to an instanton brane, because they lead to additional zero-modes that kill the amplitude. As is clear from the tables,
they are completely absent for the instantons we consider.

{\scriptsize
\begin{center}
\begin{longtable}{|c|c|c|c|c|c|c|c|c|c|c|c|c|c|c|c|c|c|c|c|c|c|c|c|c|c|c|c|c|c|c|c|c|c|c|c|c|c|c|c|}
\caption{\small Complete spectrum of a global model of type I.}\label{tab global model}\\
 \hline
   \multicolumn{1}{|l|}{Num.}
& \multicolumn{1}{l|}{Mult.}
& \multicolumn{1}{l|}{U(5)}
& \multicolumn{1}{l|}{U(1)}
& \multicolumn{1}{r|}{O(1)}
& \multicolumn{1}{l|}{O(2)}
& \multicolumn{1}{l|}{O(1)}
& \multicolumn{1}{l|}{O(1)}
& \multicolumn{1}{l|}{U(1)}
& \multicolumn{1}{l|}{O(1)}
& \multicolumn{1}{l|}{O(1)}
& \multicolumn{1}{l|}{U(3)}
& \multicolumn{1}{l|}{Inst}
& \multicolumn{1}{l|}{Chir.} \\
\hline
\endfirsthead
\multicolumn{8}{c}%
{{\bfseries \tablename\ \thetable{} {\rm-- continued from previous page}}} \\
\hline
   \multicolumn{1}{|l|}{Num.}
& \multicolumn{1}{l|}{Mult.}
& \multicolumn{1}{l|}{U(5)}
& \multicolumn{1}{l|}{U(1)}
& \multicolumn{1}{r|}{O(1)}
& \multicolumn{1}{l|}{O(2)}
& \multicolumn{1}{l|}{O(1)}
& \multicolumn{1}{l|}{O(1)}
& \multicolumn{1}{l|}{U(1)}
& \multicolumn{1}{l|}{O(1)}
& \multicolumn{1}{l|}{O(1)}
& \multicolumn{1}{l|}{U(3)}
& \multicolumn{1}{l|}{Inst}
& \multicolumn{1}{l|}{Chir.} \\ \hline
\endhead
\hline \multicolumn{16}{|l|}{{Continued on next page}} \\ \hline
\endfoot
\hline \hline
\endlastfoot
1 & 3  &  A & 0 & 0 & 0 & 0 & 0 & 0 & 0 & 0 & 0 & 0  & 3 \\
2 & 3  &  V & V$^*$ & 0 & 0 & 0 & 0 & 0 & 0 & 0 & 0 & 0  & -3 \\
3 & 3  &  V & V & 0 & 0 & 0 & 0 & 0 & 0 & 0 & 0 & 0  & -3 \\
4 & 3  &  V & 0 & V & 0 & 0 & 0 & 0 & 0 & 0 & 0 & 0  & 3 \\
5 & 12  &  0 & V & V & 0 & 0 & 0 & 0 & 0 & 0 & 0 & 0  & 0 \\
\hline 
6 & 1  &  V & 0 & 0 & 0 & 0 & 0 & 0 & 0 & 0 & 0 & V  & -1 \\
7 & 1  &  0 & 0 & V & 0 & 0 & 0 & 0 & 0 & 0 & 0 & V  & 0 \\
\hline 
8& 1  &  V & 0 & 0 & V & 0 & 0 & 0 & 0 & 0 & 0 & 0  & 1 \\
9 & 1  &  V & 0 & 0 & 0 & 0 & 0 & V$^*$ & 0 & 0 & 0 & 0  & 1 \\
10 & 1  &  V & 0 & 0 & 0 & 0 & 0 & 0 & V & 0 & 0 & 0  & -1 \\
11 & 1  &  V & 0 & 0 & 0 & 0 & 0 & 0 & 0 & V & 0 & 0  & 1 \\
12& 1  &  V & 0 & 0 & 0 & 0 & 0 & 0 & 0 & 0 & V$^*$ & 0  & -1 \\
13 & 1  &  0 & V & 0 & 0 & 0 & 0 & V$^*$ & 0 & 0 & 0 & 0  & 1 \\
14 & 1  &  0 & V & 0 & 0 & 0 & 0 & V & 0 & 0 & 0 & 0  & -1 \\
15 & 1  &  0 & 0 & V & 0 & 0 & 0 & 0 & 0 & 0 & V & 0  & -1 \\
\hline 
16 & 2  &  V & 0 & 0 & 0 & 0 & 0 & V & 0 & 0 & 0 & 0  & 0 \\
17 & 4  &  V & 0 & 0 & 0 & 0 & 0 & 0 & 0 & 0 & V & 0  & 0 \\
18 & 4  &  0 & V & 0 & V & 0 & 0 & 0 & 0 & 0 & 0 & 0  & 0 \\
19 & 6  &  0 & V & 0 & 0 & V & 0 & 0 & 0 & 0 & 0 & 0  & 0 \\
20 & 2  &  0 & V & 0 & 0 & 0 & V & 0 & 0 & 0 & 0 & 0  & 0 \\
21 & 2  &  0 & 0 & V & V & 0 & 0 & 0 & 0 & 0 & 0 & 0  & 0 \\
22 & 2  &  0 & 0 & V & 0 & V & 0 & 0 & 0 & 0 & 0 & 0  & 0 \\
23 & 4  &  0 & 0 & V & 0 & 0 & 0 & 0 & V & 0 & 0 & 0  & 0 \\
\hline 
24 & 3  & Ad & 0 & 0 & 0 & 0 & 0 & 0 & 0 & 0 & 0 & 0  & 0 \\
25 & 2  &  S & 0 & 0 & 0 & 0 & 0 & 0 & 0 & 0 & 0 & 0  & 0 \\
26 & 7  &  0 &  Ad & 0 & 0 & 0 & 0 & 0 & 0 & 0 & 0 & 0  & 0 \\
27 & 8  &  0 & S & 0 & 0 & 0 & 0 & 0 & 0 & 0 & 0 & 0  & 0 \\
28 & 8  &  0 & A & 0 & 0 & 0 & 0 & 0 & 0 & 0 & 0 & 0  & 0 \\
29 & 5  &  0 & 0 & A & 0 & 0 & 0 & 0 & 0 & 0 & 0 & 0  & 0 \\
30 & 4  &  0 & 0 & S & 0 & 0 & 0 & 0 & 0 & 0 & 0 & 0  & 0 \\
\hline
31 & 1  &  0 & 0 & 0 & V & 0 & 0 & V & 0 & 0 & 0 & 0  & 1 \\
32 & 1  &  0 & 0 & 0 & 0 & V & 0 & V & 0 & 0 & 0 & 0  & -1 \\
33 & 1  &  0 & 0 & 0 & 0 & 0 & V & V & 0 & 0 & 0 & 0  & -1 \\
34 & 1  &  0 & 0 & 0 & 0 & 0 & 0 & S & 0 & 0 & 0 & 0  & 1 \\
35 & 1  &  0 & 0 & 0 & 0 & 0 & 0 & A & 0 & 0 & 0 & 0  & -1 \\
36 & 1  &  0 & 0 & 0 & 0 & 0 & 0 & V & V & 0 & 0 & 0  & 1 \\
37 & 1  &  0 & 0 & 0 & 0 & 0 & 0 & V & 0 & V & 0 & 0  & 1 \\
38 & 1  &  0 & 0 & 0 & 0 & 0 & 0 & V & 0 & 0 & V & 0  & -1 \\
39 & 1  &  0 & 0 & 0 & 0 & 0 & 0 & 0 & V & 0 & V & 0  & 1 \\
40 & 2  &  0 & 0 & 0 & 0 & 0 & 0 & 0 & 0 & V & V & 0  & 2 \\
41 & 1  &  0 & 0 & 0 & 0 & 0 & 0 & 0 & 0 & 0 & S & 0  & -1 \\
42 & 1  &  0 & 0 & 0 & 0 & 0 & 0 & 0 & 0 & 0 & A & 0  & -1 \\
\hline
43 & 1  &  0 & 0 & 0 & S & 0 & 0 & 0 & 0 & 0 & 0 & 0  & 0 \\
44 & 1  &  0 & 0 & 0 & V & V & 0 & 0 & 0 & 0 & 0 & 0  & 0 \\
45 & 1  &  0 & 0 & 0 & 0 & V & V & 0 & 0 & 0 & 0 & 0  & 0 \\
46 & 1  &  0 & 0 & 0 & 0 & S & 0 & 0 & 0 & 0 & 0 & 0  & 0 \\
47 & 1  &  0 & 0 & 0 & 0 & V & V & 0 & 0 & 0 & 0 & 0  & 0 \\
48 & 2  &  0 & 0 & 0 & 0 & V & 0 & 0 & V & 0 & 0 & 0  & 0 \\
49 & 2  &  0 & 0 & 0 & 0 & V & 0 & 0 & 0 & V & 0 & 0  & 0 \\
50 & 1  &  0 & 0 & 0 & 0 & 0 & S & 0 & 0 & 0 & 0 & 0  & 0 \\
51 & 1  &  0 & 0 & 0 & 0 & 0 & V & 0 & V & 0 & 0 & 0  & 0 \\
52 & 1  &  0 & 0 & 0 & 0 & 0 & V & 0 & 0 & V & 0 & 0  & 0 \\
53 & 1  &  0 & 0 & 0 & 0 & 0 & 0 &  Ad & 0 & 0 & 0 & 0  & 0 \\
54 & 1  &  0 & 0 & 0 & 0 & 0 & 0 & 0 & A & 0 & 0 & 0  & 0 \\
\end{longtable}
\end{center}}

\subsection{Gepner Orientifold of Type II}

The difference to the six solutions above is that the hidden gauge symmetry is slightly different. The whole gauge symmetry is
$\mathbf{U( 5 )} \times \mathbf{U( 1 )} \times \mathbf{ O( 1 )} \times U( 1 ) \times O( 2 ) \times O( 1 ) \times O( 2 ) \times O( 1 ) \times O( 1 ) \times O( 1 ) \times U( 3 )$, where again the first three  gauge groups denote the visible sector. We find 12 different solutions which have again the same spectrum in the visible sector but different massless spectrum in the hidden sector. The Gepner orientifold is the same as above. Below we display the spectrum of one representative.

Again  we  divide the table into the different  segments: the standard model fields (1-5), where the neutrinos arise from the non-chiral sector displayed in line 5, the instanton zero modes (6-7), chiral observable-hidden matter (8-16), non-chiral observable-hidden matter (17-24), non-chiral observable rank two tensors (25-31), chiral matter within the hidden sector (32-41), and non-chiral matter within the hidden sector (42-59).

{\scriptsize
\begin{center}
\begin{longtable}{|c|c|c|c|c|c|c|c|c|c|c|c|c|c|c|c|c|c|c|c|c|c|c|c|c|c|c|c|c|c|c|c|c|c|c|c|c|c|c|c|}
\caption{ Complete spectrum of a global model of type II.}\label{tab global model 2}\\
 \hline
   \multicolumn{1}{|l|}{Num.}
& \multicolumn{1}{l|}{Mult.}
& \multicolumn{1}{l|}{U(5)}
& \multicolumn{1}{l|}{U(1)}
& \multicolumn{1}{r|}{O(1)}
& \multicolumn{1}{l|}{U(1)}
& \multicolumn{1}{l|}{O(2)}
& \multicolumn{1}{l|}{O(1)}
& \multicolumn{1}{l|}{O(2)}
& \multicolumn{1}{l|}{O(1)}
& \multicolumn{1}{l|}{O(1)}
& \multicolumn{1}{l|}{O(1)}
& \multicolumn{1}{l|}{U(3)}
& \multicolumn{1}{l|}{Inst.}
& \multicolumn{1}{l|}{Chir.} \\ \hline
\endfirsthead
\multicolumn{8}{c}%
{{\bfseries \tablename\ \thetable{} {\rm-- continued from previous page}}} \\
\hline
   \multicolumn{1}{|l|}{Num.}
& \multicolumn{1}{l|}{Mult.}
& \multicolumn{1}{l|}{U(5)}
& \multicolumn{1}{l|}{U(1)}
& \multicolumn{1}{r|}{O(1)}
& \multicolumn{1}{l|}{U(1)}
& \multicolumn{1}{l|}{O(2)}
& \multicolumn{1}{l|}{O(1)}
& \multicolumn{1}{l|}{O(2)}
& \multicolumn{1}{l|}{O(1)}
& \multicolumn{1}{l|}{O(1)}
& \multicolumn{1}{l|}{O(1)}
& \multicolumn{1}{l|}{U(3)}
& \multicolumn{1}{l|}{Inst.}
& \multicolumn{1}{l|}{Chir.} \\ \hline
\endhead
\hline \multicolumn{16}{|l|}{{Continued on next page}} \\ \hline
\endfoot
\hline \hline
\endlastfoot
1&3  &  A & 0 & 0 & 0 & 0 & 0 & 0 & 0 & 0 & 0 & 0 & 0  & 3   \\
2&3  &  V & V$^*$ & 0 & 0 & 0 & 0 & 0 & 0 & 0 & 0 & 0 & 0  & -3  \\
3&3  &  V & V & 0 & 0 & 0 & 0 & 0 & 0 & 0 & 0 & 0 & 0  & -3 \\
4&3  &  V & 0 & V & 0 & 0 & 0 & 0 & 0 & 0 & 0 & 0 & 0  & 3  \\
5&6  &  0 & V & V & 0 & 0 & 0 & 0 & 0 & 0 & 0 & 0 & 0  & 0  \\
\hline
6&1  &  V & 0 & 0 & 0 & 0 & 0 & 0 & 0 & 0 & 0 & 0 & V  & -1 \\
7&1  &  0 & 0 & V & 0 & 0 & 0 & 0 & 0 & 0 & 0 & 0 & V  & 0  \\
\hline
8&1  &  V & 0 & 0 & 0 & V & 0 & 0 & 0 & 0 & 0 & 0 & 0  & 1 \\
9&1  &  V & 0 & 0 & 0 & 0 & 0 & V & 0 & 0 & 0 & 0 & 0  & 1 \\
10&1  &  V & 0 & 0 & 0 & 0 & 0 & 0 & V & 0 & 0 & 0 & 0  & -1 \\
11&1  &  V & 0 & 0 & 0 & 0 & 0 & 0 & 0 & V & 0 & 0 & 0  & -1 \\
12&1  &  V & 0 & 0 & 0 & 0 & 0 & 0 & 0 & 0 & V & 0 & 0  & 1 \\
13&1  &  V & 0 & 0 & 0 & 0 & 0 & 0 & 0 & 0 & 0 & V$^*$ & 0  & -1 \\
14&1  &  0 & V & 0 & V & 0 & 0 & 0 & 0 & 0 & 0 & 0 & 0  & 1 \\
15&1  &  0 & V & 0 & V$^*$ & 0 & 0 & 0 & 0 & 0 & 0 & 0 & 0  & -1 \\
16&1  &  0 & 0 & V & V & 0 & 0 & 0 & 0 & 0 & 0 & 0 & 0  & 1 \\
\hline
17&4  &  V & 0 & 0 & 0 & 0 & 0 & 0 & 0 & 0 & 0 & V & 0  & 0 \\
18&4  &  0 & V & 0 & 0 & V & 0 & 0 & 0 & 0 & 0 & 0 & 0  & 0 \\
19&2  &  0 & V & 0 & 0 & 0 & V & 0 & 0 & 0 & 0 & 0 & 0  & 0 \\
20&6  &  0 & V & 0 & 0 & 0 & 0 & V & 0 & 0 & 0 & 0 & 0  & 0 \\
21&1  &  0 & 0 & V & 0 & V & 0 & 0 & 0 & 0 & 0 & 0 & 0  & 0  \\
22&1  &  0 & 0 & V & 0 & 0 & 0 & V & 0 & 0 & 0 & 0 & 0  & 0 \\
23&3  &  0 & 0 & V & 0 & 0 & 0 & 0 & V & 0 & 0 & 0 & 0  & 0 \\
24&4  &  0 & 0 & V & 0 & 0 & 0 & 0 & 0 & V & 0 & 0 & 0  & 0 \\
\hline
25&3  &Ad & 0 & 0 & 0 & 0 & 0 & 0 & 0 & 0 & 0 & 0 & 0  & 0 \\
26&2  &  S & 0 & 0 & 0 & 0 & 0 & 0 & 0 & 0 & 0 & 0 & 0  & 0 \\
27&7  &  0 &  Ad & 0 & 0 & 0 & 0 & 0 & 0 & 0 & 0 & 0 & 0  & 0 \\
28&8  &  0 & S & 0 & 0 & 0 & 0 & 0 & 0 & 0 & 0 & 0 & 0  & 0 \\
29&8  &  0 & A & 0 & 0 & 0 & 0 & 0 & 0 & 0 & 0 & 0 & 0  & 0 \\
30&1  &  0 & 0 & S & 0 & 0 & 0 & 0 & 0 & 0 & 0 & 0 & 0  & 0 \\
31&1  &  0 & 0 & A & 0 & 0 & 0 & 0 & 0 & 0 & 0 & 0 & 0  & 0 \\
\hline
32&1  &  0 & 0 & 0 & S & 0 & 0 & 0 & 0 & 0 & 0 & 0 & 0  & -1 \\
33&1  &  0 & 0 & 0 & A & 0 & 0 & 0 & 0 & 0 & 0 & 0 & 0  & 1 \\
34&1  &  0 & 0 & 0 & V & V & 0 & 0 & 0 & 0 & 0 & 0 & 0  & 1 \\
35&1  &  0 & 0 & 0 & V & 0 & V & 0 & 0 & 0 & 0 & 0 & 0  & 1 \\
36&1  &  0 & 0 & 0 & V & 0 & 0 & V & 0 & 0 & 0 & 0 & 0  & 1 \\
37&1  &  0 & 0 & 0 & 0 & 0 & 0 & V & 0 & 0 & 0 & V & 0  & -1 \\
38&1  &  0 & 0 & 0 & 0 & 0 & 0 & 0 & V & 0 & 0 & V & 0  & 1 \\
39&2  &  0 & 0 & 0 & 0 & 0 & 0 & 0 & 0 & 0 & V & V & 0  & 2 \\
40&1  &  0 & 0 & 0 & 0 & 0 & 0 & 0 & 0 & 0 & 0 & S & 0  & -1 \\
41&1  &  0 & 0 & 0 & 0 & 0 & 0 & 0 & 0 & 0 & 0 & A & 0  & -1 \\

\hline
42&1  &  0 & 0 & 0 &  Ad & 0 & 0 & 0 & 0 & 0 & 0 & 0 & 0  & 0 \\
43&2  &  0 & 0 & 0 & V & 0 & 0 & 0 & 0 & 0 & 0 & V$^*$ & 0  & 0 \\
44&1  &  0 & 0 & 0 & 0 & S & 0 & 0 & 0 & 0 & 0 & 0 & 0  & 0 \\
45&1  &  0 & 0 & 0 & 0 & V & V & 0 & 0 & 0 & 0 & 0 & 0  & 0 \\
46&1  &  0 & 0 & 0 & 0 & V & 0 & V & 0 & 0 & 0 & 0 & 0  & 0 \\
47&1  &  0 & 0 & 0 & 0 & 0 & S & 0 & 0 & 0 & 0 & 0 & 0  & 0 \\
48&1  &  0 & 0 & 0 & 0 & 0 & V & V & 0 & 0 & 0 & 0 & 0  & 0 \\
49&1  &  0 & 0 & 0 & 0 & 0 & V & 0 & V & 0 & 0 & 0 & 0  & 0 \\
50&1  &  0 & 0 & 0 & 0 & 0 & V & 0 & 0 & V & 0 & 0 & 0  & 0 \\
51&1  &  0 & 0 & 0 & 0 & 0 & V & 0 & 0 & 0 & V & 0 & 0  & 0 \\
52&1  &  0 & 0 & 0 & 0 & 0 & 0 & S & 0 & 0 & 0 & 0 & 0  & 0 \\
53&3  &  0 & 0 & 0 & 0 & 0 & 0 & V & V & 0 & 0 & 0 & 0  & 0 \\
54&3  &  0 & 0 & 0 & 0 & 0 & 0 & V & 0 & V & 0 & 0 & 0  & 0 \\
55&1  &  0 & 0 & 0 & 0 & 0 & 0 & V & 0 & 0 & V & 0 & 0  & 0 \\
56&1  &  0 & 0 & 0 & 0 & 0 & 0 & 0 & A & 0 & 0 & 0 & 0  & 0 \\
57&2  &  0 & 0 & 0 & 0 & 0 & 0 & 0 & V & V & 0 & 0 & 0  & 0 \\
58&2  &  0 & 0 & 0 & 0 & 0 & 0 & 0 & 0 & A & 0 & 0 & 0  & 0 \\
59&1  &  0 & 0 & 0 & 0 & 0 & 0 & 0 & 0 & V & V & 0 & 0  & 0 \\
\end{longtable}
\end{center}}

\clearpage \nocite{*}

\begingroup\raggedright\endgroup


\begin{thebibliography}{10}

\bibitem{Kiritsis:2009sf}
E.~Kiritsis, M.~Lennek, and B.~Schellekens, {\it {SU(5) orientifolds, Yukawa
  couplings, Stringy Instantons and Proton Decay}},  {\em Nucl. Phys.} {\bf
  B829} (2010) 298--324, [\href{http://xxx.lanl.gov/abs/0909.0271}{{\tt
  arXiv:0909.0271}}].

\bibitem{Blumenhagen:2005mu}
R.~Blumenhagen, M.~Cveti{\v c}, P.~Langacker, and G.~Shiu, {\it {Toward
  realistic intersecting D-brane models}},  {\em Ann. Rev. Nucl. Part. Sci.}
  {\bf 55} (2005) 71--139, [\href{http://xxx.lanl.gov/abs/hep-th/0502005}{{\tt
  hep-th/0502005}}].

\bibitem{Blumenhagen:2006ci}
R.~Blumenhagen, B.~K{\"o}rs, D.~L{\"u}st, and S.~Stieberger, {\it
  {Four-dimensional String Compactifications with D-Branes, Orientifolds and
  Fluxes}},  {\em Phys. Rept.} {\bf 445} (2007) 1--193,
  [\href{http://xxx.lanl.gov/abs/hep-th/0610327}{{\tt hep-th/0610327}}].

\bibitem{Marchesano:2007de}
F.~Marchesano, {\it {Progress in D-brane model building}},  {\em Fortsch.
  Phys.} {\bf 55} (2007) 491--518,
  [\href{http://xxx.lanl.gov/abs/hep-th/0702094}{{\tt hep-th/0702094}}].

\bibitem{Cvetic:2002pj}
M.~Cveti{\v c}, I.~Papadimitriou, and G.~Shiu, {\it {Supersymmetric three
  family SU(5) grand unified models from type IIA orientifolds with
  intersecting D6-branes}},  {\em Nucl. Phys.} {\bf B659} (2003) 193--223,
  [\href{http://xxx.lanl.gov/abs/hep-th/0212177}{{\tt hep-th/0212177}}].

\bibitem{Chen:2005aba}
C.~M. Chen, G.~V. Kraniotis, V.~E. Mayes, D.~V. Nanopoulos, and J.~W. Walker,
  {\it {A supersymmetric flipped SU(5) intersecting brane world}},  {\em Phys.
  Lett.} {\bf B611} (2005) 156--166,
  [\href{http://xxx.lanl.gov/abs/hep-th/0501182}{{\tt hep-th/0501182}}].

\bibitem{Chen:2005mm}
C.~M. Chen, G.~V. Kraniotis, V.~E. Mayes, D.~V. Nanopoulos, and J.~W. Walker,
  {\it {A K-theory anomaly free supersymmetric flipped SU(5) model from
  intersecting branes}},  {\em Phys. Lett.} {\bf B625} (2005) 96--105,
  [\href{http://xxx.lanl.gov/abs/hep-th/0507232}{{\tt hep-th/0507232}}].

\bibitem{Chen:2005cf}
C.-M. Chen, V.~E. Mayes, and D.~V. Nanopoulos, {\it {Flipped SU(5) from
  D-branes with type IIB fluxes}},  {\em Phys. Lett.} {\bf B633} (2006)
  618--626, [\href{http://xxx.lanl.gov/abs/hep-th/0511135}{{\tt
  hep-th/0511135}}].

\bibitem{Gmeiner:2006vb}
F.~Gmeiner and M.~Stein, {\it {Statistics of SU(5) D-brane models on a type II
  orientifold}},  {\em Phys. Rev.} {\bf D73} (2006) 126008,
  [\href{http://xxx.lanl.gov/abs/hep-th/0603019}{{\tt hep-th/0603019}}].

\bibitem{Floratos:2006hs}
E.~Floratos and C.~Kokorelis, {\it {MSSM GUT string vacua, split supersymmetry
  and fluxes}},  \href{http://xxx.lanl.gov/abs/hep-th/0607217}{{\tt
  hep-th/0607217}}.

\bibitem{Cvetic:2006by}
M.~Cveti{\v c} and P.~Langacker, {\it {New grand unified models with
  intersecting D6-branes, neutrino masses, and flipped SU(5)}},  {\em Nucl.
  Phys.} {\bf B776} (2007) 118--137,
  [\href{http://xxx.lanl.gov/abs/hep-th/0607238}{{\tt hep-th/0607238}}].

\bibitem{Antoniadis:2007jq}
I.~Antoniadis, A.~Kumar, and B.~Panda, {\it {Supersymmetric SU(5) GUT with
  Stabilized Moduli}},  {\em Nucl. Phys.} {\bf B795} (2008) 69--104,
  [\href{http://xxx.lanl.gov/abs/0709.2799}{{\tt arXiv:0709.2799}}].

\bibitem{Blumenhagen:2008zz}
R.~Blumenhagen, V.~Braun, T.~W. Grimm, and T.~Weigand, {\it {GUTs in Type IIB
  Orientifold Compactifications}},  {\em Nucl. Phys.} {\bf B815} (2009) 1--94,
  [\href{http://xxx.lanl.gov/abs/0811.2936}{{\tt arXiv:0811.2936}}].

\bibitem{Berenstein:2006aj}
D.~Berenstein, {\it {Branes vs. GUTS: Challenges for string inspired
  phenomenology}},  \href{http://xxx.lanl.gov/abs/hep-th/0603103}{{\tt
  hep-th/0603103}}.

\bibitem{Blumenhagen:2006xt}
R.~Blumenhagen, M.~Cveti{\v c}, and T.~Weigand, {\it Spacetime instanton
  corrections in 4d string vacua - the seesaw mechanism for d-brane models},
  {\em Nucl. Phys.} {\bf B771} (2007) 113--142,
  [\href{http://xxx.lanl.gov/abs/hep-th/0609191}{{\tt hep-th/0609191}}].

\bibitem{Ibanez:2006da}
L.~E. Ib{\'a}{\~n}ez and A.~M. Uranga, {\it {Neutrino Majorana masses from
  string theory instanton effects}},  {\em JHEP} {\bf 03} (2007) 052,
  [\href{http://xxx.lanl.gov/abs/hep-th/0609213}{{\tt hep-th/0609213}}].

\bibitem{Florea:2006si}
B.~Florea, S.~Kachru, J.~McGreevy, and N.~Saulina, {\it {Stringy Instantons and
  Quiver Gauge Theories}},  {\em JHEP} {\bf 05} (2007) 024,
  [\href{http://xxx.lanl.gov/abs/hep-th/0610003}{{\tt hep-th/0610003}}].

\bibitem{Blumenhagen:2009qh}
R.~Blumenhagen, M.~Cveti{\v c}, S.~Kachru, and T.~Weigand, {\it {{\small
  D}-Brane Instantons in Type {II} Orientifolds}},  {\em Ann. Rev. Nucl. Part.
  Sci.} {\bf 59} (2009) 269--296,
  [\href{http://xxx.lanl.gov/abs/0902.3251}{{\tt arXiv:0902.3251}}].

\bibitem{Bianchi:2009ij}
M.~Bianchi and M.~Samsonyan, {\it {Notes on unoriented D-brane instantons}},
  {\em Int. J. Mod. Phys.} {\bf A24} (2009) 5737--5763,
  [\href{http://xxx.lanl.gov/abs/0909.2173}{{\tt arXiv:0909.2173}}].

\bibitem{Argurio:2007qk}
R.~Argurio, M.~Bertolini, S.~Franco, and S.~Kachru, {\it {Metastable vacua and
  D-branes at the conifold}},  {\em JHEP} {\bf 06} (2007) 017,
  [\href{http://xxx.lanl.gov/abs/hep-th/0703236}{{\tt hep-th/0703236}}].

\bibitem{Argurio:2007vq}
R.~Argurio, M.~Bertolini, G.~Ferretti, A.~Lerda, and C.~Petersson, {\it Stringy
  instantons at orbifold singularities},  {\em JHEP} {\bf 06} (2007) 067,
  [\href{http://xxx.lanl.gov/abs/0704.0262}{{\tt 0704.0262}}].

\bibitem{Bianchi:2007wy}
M.~Bianchi, F.~Fucito, and J.~F. Morales, {\it D-brane instantons on the t6/z3
  orientifold},  {\em JHEP} {\bf 07} (2007) 038,
  [\href{http://xxx.lanl.gov/abs/0704.0784}{{\tt 0704.0784}}].

\bibitem{Ibanez:2007rs}
L.~E. Ib{\'a}{\~n}ez, A.~N. Schellekens, and A.~M. Uranga, {\it Instanton
  induced neutrino majorana masses in cft orientifolds with mssm-like spectra},
   {\em JHEP} {\bf 06} (2007) 011,
  [\href{http://xxx.lanl.gov/abs/0704.1079}{{\tt 0704.1079}}].

\bibitem{Blumenhagen:2007bn}
R.~Blumenhagen, M.~Cveti{\v c}, R.~Richter, and T.~Weigand, {\it {Lifting
  D-Instanton Zero Modes by Recombination and Background Fluxes}},  {\em JHEP}
  {\bf 10} (2007) 098, [\href{http://xxx.lanl.gov/abs/0708.0403}{{\tt
  arXiv:0708.0403}}].

\bibitem{GarciaEtxebarria:2007zv}
I.~Garc{\'i}a-Etxebarria and A.~M. Uranga, {\it {Non-perturbative
  superpotentials across lines of marginal stability}},  {\em JHEP} {\bf 01}
  (2008) 033, [\href{http://xxx.lanl.gov/abs/0711.1430}{{\tt
  arXiv:0711.1430}}].

\bibitem{Cvetic:2008ws}
M.~Cveti{\v c}, R.~Richter, and T.~Weigand, {\it {(Non-)BPS bound states and
  D-brane instantons}},  {\em JHEP} {\bf 07} (2008) 012,
  [\href{http://xxx.lanl.gov/abs/0803.2513}{{\tt arXiv:0803.2513}}].

\bibitem{GarciaEtxebarria:2008pi}
I.~Garc{\'i}a-Etxebarria, F.~Marchesano, and A.~M. Uranga, {\it
  {Non-perturbative F-terms across lines of BPS stability}},  {\em JHEP} {\bf
  07} (2008) 028, [\href{http://xxx.lanl.gov/abs/0805.0713}{{\tt
  arXiv:0805.0713}}].

\bibitem{Aganagic:2007py}
M.~Aganagic, C.~Beem, and S.~Kachru, {\it {Geometric Transitions and Dynamical
  SUSY Breaking}},  {\em Nucl. Phys.} {\bf B796} (2008) 1--24,
  [\href{http://xxx.lanl.gov/abs/0709.4277}{{\tt arXiv:0709.4277}}].

\bibitem{Petersson:2007sc}
C.~Petersson, {\it {Superpotentials From Stringy Instantons Without
  Orientifolds}},  {\em JHEP} {\bf 05} (2008) 078,
  [\href{http://xxx.lanl.gov/abs/0711.1837}{{\tt arXiv:0711.1837}}].

\bibitem{GarciaEtxebarria:2008iw}
I.~Garc{\'i}a-Etxebarria, {\it {D-brane instantons and matrix models}},  {\em
  JHEP} {\bf 07} (2009) 017, [\href{http://xxx.lanl.gov/abs/0810.1482}{{\tt
  arXiv:0810.1482}}].

\bibitem{Ferretti:2009tz}
G.~Ferretti and C.~Petersson, {\it {Non-Perturbative Effects on a Fractional
  D3-Brane}},  {\em JHEP} {\bf 03} (2009) 040,
  [\href{http://xxx.lanl.gov/abs/0901.1182}{{\tt arXiv:0901.1182}}].

\bibitem{Cvetic:2009ez}
M.~Cveti{\v c}, J.~Halverson, and R.~Richter, {\it {Mass Hierarchies from MSSM
  Orientifold Compactifications}},  {\em JHEP} {\bf 07} (2010) 005,
  [\href{http://xxx.lanl.gov/abs/0909.4292}{{\tt arXiv:0909.4292}}].

\bibitem{Blumenhagen:2007zk}
R.~Blumenhagen, M.~Cveti{\v c}, D.~L{\"u}st, R.~Richter, and T.~Weigand, {\it
  {Non-perturbative Yukawa Couplings from String Instantons}},  {\em Phys. Rev.
  Lett.} {\bf 100} (2008) 061602,
  [\href{http://xxx.lanl.gov/abs/0707.1871}{{\tt arXiv:0707.1871}}].

\bibitem{Cvetic:2007ku}
M.~Cveti{\v c}, R.~Richter, and T.~Weigand, {\it {Computation of D-brane
  instanton induced superpotential couplings - Majorana masses from string
  theory}},  {\em Phys. Rev.} {\bf D76} (2007) 086002,
  [\href{http://xxx.lanl.gov/abs/hep-th/0703028}{{\tt hep-th/0703028}}].

\bibitem{Antusch:2007jd}
S.~Antusch, L.~E. Ib{\'a}{\~n}ez, and T.~Macri, {\it {Neutrino Masses and
  Mixings from String Theory Instantons}},  {\em JHEP} {\bf 09} (2007) 087,
  [\href{http://xxx.lanl.gov/abs/0706.2132}{{\tt arXiv:0706.2132}}].

\bibitem{Cvetic:2007qj}
M.~Cveti{\v c} and T.~Weigand, {\it {Hierarchies from D-brane instantons in
  globally defined Calabi-Yau Orientifolds}},  {\em Phys. Rev. Lett.} {\bf 100}
  (2008) 251601, [\href{http://xxx.lanl.gov/abs/0711.0209}{{\tt
  arXiv:0711.0209}}].

\bibitem{Ibanez:2008my}
L.~E. Ib{\'a}{\~n}ez and R.~Richter, {\it {Stringy Instantons and Yukawa
  Couplings in MSSM-like Orientifold Models}},  {\em JHEP} {\bf 03} (2009) 090,
  [\href{http://xxx.lanl.gov/abs/0811.1583}{{\tt arXiv:0811.1583}}].

\bibitem{Cvetic:2008hi}
M.~Cveti{\v c} and P.~Langacker, {\it {D-Instanton Generated Dirac Neutrino
  Masses}},  {\em Phys. Rev.} {\bf D78} (2008) 066012,
  [\href{http://xxx.lanl.gov/abs/0803.2876}{{\tt arXiv:0803.2876}}].

\bibitem{Angelantonj:1996mw}
C.~Angelantonj, M.~Bianchi, G.~Pradisi, A.~Sagnotti, and Y.~S. Stanev, {\it
  {Comments on Gepner models and type I vacua in string theory}},  {\em Phys.
  Lett.} {\bf B387} (1996) 743--749,
  [\href{http://xxx.lanl.gov/abs/hep-th/9607229}{{\tt hep-th/9607229}}].

\bibitem{Blumenhagen:1998tj}
R.~Blumenhagen and A.~Wisskirchen, {\it {Spectra of 4D, N = 1 type I string
  vacua on non-toroidal CY threefolds}},  {\em Phys. Lett.} {\bf B438} (1998)
  52--60, [\href{http://xxx.lanl.gov/abs/hep-th/9806131}{{\tt
  hep-th/9806131}}].

\bibitem{Blumenhagen:2003su}
R.~Blumenhagen, {\it {Supersymmetric orientifolds of Gepner models}},  {\em
  JHEP} {\bf 11} (2003) 055,
  [\href{http://xxx.lanl.gov/abs/hep-th/0310244}{{\tt hep-th/0310244}}].

\bibitem{Blumenhagen:2004cg}
R.~Blumenhagen and T.~Weigand, {\it {Chiral supersymmetric Gepner model
  orientifolds}},  {\em JHEP} {\bf 02} (2004) 041,
  [\href{http://xxx.lanl.gov/abs/hep-th/0401148}{{\tt hep-th/0401148}}].

\bibitem{Blumenhagen:2004qu}
R.~Blumenhagen and T.~Weigand, {\it {A note on partition functions of Gepner
  model orientifolds}},  {\em Phys. Lett.} {\bf B591} (2004) 161--169,
  [\href{http://xxx.lanl.gov/abs/hep-th/0403299}{{\tt hep-th/0403299}}].

\bibitem{Blumenhagen:2004hd}
R.~Blumenhagen and T.~Weigand, {\it {Chiral Gepner model orientifolds}},
  \href{http://xxx.lanl.gov/abs/hep-th/0408147}{{\tt hep-th/0408147}}.

\bibitem{Gepner:1987vz}
D.~Gepner, {\it {Exactly Solvable String Compactifications on Manifolds of
  SU(N) Holonomy}},  {\em Phys. Lett.} {\bf B199} (1987) 380--388.

\bibitem{Gepner:1987qi}
D.~Gepner, {\it {Space-Time Supersymmetry in Compactified String Theory and
  Superconformal Models}},  {\em Nucl. Phys.} {\bf B296} (1988) 757.

\bibitem{Eguchi:1988vra}
T.~Eguchi, H.~Ooguri, A.~Taormina, and S.-K. Yang, {\it {Superconformal
  Algebras and String Compactification on Manifolds with SU(N) Holonomy}},
  {\em Nucl. Phys.} {\bf B315} (1989) 193.

\bibitem{Dijkstra:2004ym}
T.~P.~T. Dijkstra, L.~R. Huiszoon, and A.~N. Schellekens, {\it {Chiral
  supersymmetric standard model spectra from orientifolds of Gepner models}},
  {\em Phys. Lett.} {\bf B609} (2005) 408--417,
  [\href{http://xxx.lanl.gov/abs/hep-th/0403196}{{\tt hep-th/0403196}}].

\bibitem{Dijkstra:2004cc}
T.~P.~T. Dijkstra, L.~R. Huiszoon, and A.~N. Schellekens, {\it {Supersymmetric
  Standard Model Spectra from RCFT orientifolds}},  {\em Nucl. Phys.} {\bf
  B710} (2005) 3--57, [\href{http://xxx.lanl.gov/abs/hep-th/0411129}{{\tt
  hep-th/0411129}}].

\bibitem{Anastasopoulos:2006da}
P.~Anastasopoulos, T.~P.~T. Dijkstra, E.~Kiritsis, and A.~N. Schellekens, {\it
  {Orientifolds, hypercharge embeddings and the standard model}},  {\em Nucl.
  Phys.} {\bf B759} (2006) 83--146,
  [\href{http://xxx.lanl.gov/abs/hep-th/0605226}{{\tt hep-th/0605226}}].

\bibitem{Kiritsis:2008ry}
E.~Kiritsis, B.~Schellekens, and M.~Tsulaia, {\it {Discriminating MSSM families
  in (free-field) Gepner Orientifolds}},  {\em JHEP} {\bf 10} (2008) 106,
  [\href{http://xxx.lanl.gov/abs/0809.0083}{{\tt arXiv:0809.0083}}].

\bibitem{Antoniadis:1987dx}
I.~Antoniadis, J.~R. Ellis, J.~S. Hagelin, and D.~V. Nanopoulos, {\it
  {Supersymmetric Flipped SU(5) Revitalized}},  {\em Phys. Lett.} {\bf B194}
  (1987) 231.

\bibitem{Cvetic:2009yh}
M.~Cveti{\v c}, J.~Halverson, and R.~Richter, {\it {Realistic Yukawa structures
  from orientifold compactifications}},  {\em JHEP} {\bf 12} (2009) 063,
  [\href{http://xxx.lanl.gov/abs/0905.3379}{{\tt arXiv:0905.3379}}].

\bibitem{Bianchi:2000de}
M.~Bianchi and J.~F. Morales, {\it {Anomalies and tadpoles}},  {\em JHEP} {\bf
  03} (2000) 030, [\href{http://xxx.lanl.gov/abs/hep-th/0002149}{{\tt
  hep-th/0002149}}].

\bibitem{Leontaris:2009ci}
G.~K. Leontaris, {\it {Instanton induced charged fermion and neutrino masses in
  a minimal Standard Model scenario from intersecting D- branes}},  {\em Int.
  J. Mod. Phys.} {\bf A24} (2009) 6035--6049,
  [\href{http://xxx.lanl.gov/abs/0903.3691}{{\tt arXiv:0903.3691}}].

\bibitem{Anastasopoulos:2009mr}
P.~Anastasopoulos, E.~Kiritsis, and A.~Lionetto, {\it {On mass hierarchies in
  orientifold vacua}},  {\em JHEP} {\bf 08} (2009) 026,
  [\href{http://xxx.lanl.gov/abs/0905.3044}{{\tt arXiv:0905.3044}}].

\bibitem{Cvetic:2010mm}
M.~Cveti{\v c}, J.~Halverson, P.~Langacker, and R.~Richter, {\it {The Weinberg
  Operator and a Lower String Scale in Orientifold Compactifications}},
  \href{http://xxx.lanl.gov/abs/1001.3148}{{\tt arXiv:1001.3148}}.

\bibitem{Anastasopoulos:2010ca}
P.~Anastasopoulos, G.~K. Leontaris, and N.~D. Vlachos, {\it {Phenomenological
  analysis of D-brane Pati-Salam vacua}},  {\em JHEP} {\bf 05} (2010) 011,
  [\href{http://xxx.lanl.gov/abs/1002.2937}{{\tt arXiv:1002.2937}}].

\bibitem{Cvetic:2010dz}
M.~Cveti{\v c}, J.~Halverson, and P.~Langacker, {\it {Singlet Extensions of the
  MSSM in the Quiver Landscape}},  {\em JHEP} {\bf 09} (2010) 076,
  [\href{http://xxx.lanl.gov/abs/1006.3341}{{\tt arXiv:1006.3341}}].

\bibitem{Blumenhagen:2010dt}
R.~Blumenhagen, A.~Deser, and D.~L{\"u}st, {\it {FCNC Processes from D-brane
  Instantons}},  \href{http://xxx.lanl.gov/abs/1007.4770}{{\tt
  arXiv:1007.4770}}.

\bibitem{Fucito:2010dk}
F.~Fucito, A.~Lionetto, J.~F. Morales, and R.~Richter, {\it {Dynamical
  Supersymmetry Breaking in Intersecting Brane Models}},
  \href{http://xxx.lanl.gov/abs/1007.5449}{{\tt arXiv:1007.5449}}.

\bibitem{Antoniadis:2000ena}
I.~Antoniadis, E.~Kiritsis, and T.~N. Tomaras, {\it {A D-brane alternative to
  unification}},  {\em Phys. Lett.} {\bf B486} (2000) 186--193,
  [\href{http://xxx.lanl.gov/abs/hep-ph/0004214}{{\tt hep-ph/0004214}}].

\bibitem{Aldazabal:2000sa}
G.~Aldazabal, L.~E. Ib{\'a}{\~n}ez, F.~Quevedo, and A.~M. Uranga, {\it
  {D-branes at singularities: A bottom-up approach to the string embedding of
  the standard model}},  {\em JHEP} {\bf 08} (2000) 002,
  [\href{http://xxx.lanl.gov/abs/hep-th/0005067}{{\tt hep-th/0005067}}].

\bibitem{Antoniadis:2001np}
I.~Antoniadis, E.~Kiritsis, and T.~Tomaras, {\it {D-brane Standard Model}},
  {\em Fortsch. Phys.} {\bf 49} (2001) 573--580,
  [\href{http://xxx.lanl.gov/abs/hep-th/0111269}{{\tt hep-th/0111269}}].

\bibitem{Aldazabal:2000cn}
G.~Aldazabal, S.~Franco, L.~E. Ib{\'a}{\~n}ez, R.~Rabadan, and A.~M. Uranga,
  {\it {Intersecting brane worlds}},  {\em JHEP} {\bf 02} (2001) 047,
  [\href{http://xxx.lanl.gov/abs/hep-ph/0011132}{{\tt hep-ph/0011132}}].

\bibitem{Antoniadis:2002cs}
I.~Antoniadis, E.~Kiritsis, and J.~Rizos, {\it {Anomalous U(1)s in type I
  superstring vacua}},  {\em Nucl. Phys.} {\bf B637} (2002) 92--118,
  [\href{http://xxx.lanl.gov/abs/hep-th/0204153}{{\tt hep-th/0204153}}].

\bibitem{Anastasopoulos:2003aj}
P.~Anastasopoulos, {\it {4D anomalous U(1)'s, their masses and their relation
  to 6D anomalies}},  {\em JHEP} {\bf 08} (2003) 005,
  [\href{http://xxx.lanl.gov/abs/hep-th/0306042}{{\tt hep-th/0306042}}].

\bibitem{Anastasopoulos:2004ga}
P.~Anastasopoulos, {\it {Anomalous U(1)s masses in non-supersymmetric open
  string vacua}},  {\em Phys. Lett.} {\bf B588} (2004) 119--126,
  [\href{http://xxx.lanl.gov/abs/hep-th/0402105}{{\tt hep-th/0402105}}].

\bibitem{Fuchs:2000cm}
J.~Fuchs, L.~R. Huiszoon, A.~N. Schellekens, C.~Schweigert, and J.~Walcher,
  {\it {Boundaries, crosscaps and simple currents}},  {\em Phys. Lett.} {\bf
  B495} (2000) 427--434, [\href{http://xxx.lanl.gov/abs/hep-th/0007174}{{\tt
  hep-th/0007174}}].

\bibitem{Pradisi:1996yd}
G.~Pradisi, A.~Sagnotti, and Y.~S. Stanev, {\it {Completeness Conditions for
  Boundary Operators in 2D Conformal Field Theory}},  {\em Phys. Lett.} {\bf
  B381} (1996) 97--104, [\href{http://xxx.lanl.gov/abs/hep-th/9603097}{{\tt
  hep-th/9603097}}].

\end{thebibliography}

\end{document}